\documentclass[reprint, amsmath,amssymb, aps, prb,  superscriptaddress]{revtex4-2}
\pdfoutput=1

\usepackage{geometry}
\usepackage{mathrsfs}
\usepackage{amsmath}
\usepackage{graphics}
\usepackage{color}
\usepackage{courier}
\usepackage{url}
\usepackage{listings}
\usepackage{gensymb}
\usepackage[table,xcdraw]{xcolor}
\usepackage[]{subfig}
\usepackage[colorlinks]{hyperref}
\usepackage{adjustbox}
\usepackage{xcolor}

\newcommand{\BaMoS}{\text{BaMo}_6\text{S}_8}
\newcommand{\EuMoS}{\text{EuMo}_6\text{S}_8}
\newcommand{\SrMoS}{\text{SrMo}_6\text{S}_8}

\newcommand{\MoS}{\text{Mo}_6\text{S}_8}

\begin{document}
\preprint{APS/123-QED}

\title{Topological States in Chevrel Phase Materials from First-principle Calculations}

\author{Shuai Zhang}
\affiliation{Beijing National Laboratory for Condensed Matter Physics, and Institute of Physics, Chinese Academy of Sciences, Beijing 100190, China}
\affiliation{University of Chinese Academy of Sciences, Beijing 100049, China}

\author{Shiyu Peng}
\affiliation{Beijing National Laboratory for Condensed Matter Physics, and Institute of Physics, Chinese Academy of Sciences, Beijing 100190, China}
\affiliation{University of Chinese Academy of Sciences, Beijing 100049, China}

\author{Xi Dai}
\affiliation{Department of Physics, Hong Kong University of Science and Technology, Clear Water Bay, Hong Kong}

\author{Hongming Weng}
\email{hmweng@iphy.ac.cn}
\affiliation{Beijing National Laboratory for Condensed Matter Physics, and Institute of Physics, Chinese Academy of Sciences, Beijing 100190, China}
\affiliation{University of Chinese Academy of Sciences, Beijing 100049, China}
\affiliation{Songshan Lake Materials Laboratory, Dongguan, Guangdong 523808, China}


\begin{abstract}
Chevrel phase materials form a family of ternary molybdenum chalcogenides with a general chemical formula $A_x$Mo$_6X_8$ ($A$ = metal elements, $X$ = chalcogen). 
The variety of $A$ atoms makes a large number of family members and leads to many tunable physical properties, such as the superconductivity, thermoelectricity and the ionic conductivity.
In this work, we have further found various nontrivial band topological states in these materials by using first-principle calculations.
The compounds having time-reversal symmetry, such as $\BaMoS$, $\SrMoS$ and $\MoS$, are topological insulators in both of the $R\bar{3}$ and $P\bar{1}$ phases, whereas $\EuMoS$ within ferromagnetic state is an axion insulator in the $R\bar{3}$ phase and a trivial one in the $P\bar{1}$ phase. 
This indicates that the change of $A$ ions can modify the chemical potential, lattice distortion, and magnetic orders, which offers a unique way to influence the topological states and other properties.
We hope this work can stimulate  further studies of Chevrel phase materials to find more intriguing phenomena, such as topological superconducting states and Majorana modes.
\end{abstract}

\maketitle

\section*{Introduction}
  The Chevrel phase (CP) is a ternary molybdenum chalcogenides compound family~\cite{Octavio_chevrel_2015}, 
  which was first discovered in 1971 by Chevrel and Sergent~\cite{CHEVREL1971515}.
  The generic chemical formula of  this family is $A_x\text{Mo}_6X_8$. In its crystal structure, 
  ${\rm Mo}_6X_8$ can be looked as a cluster and forms a three-dimensional (3D) network. In each cluster, six Mo atoms construct a octahedron and chalcogen atoms ($X$ = S, Se) form a distorted pyramid ligand field around each Mo.
  $A$ fills in this cluster network with $x$ varying from 0 to 4~\cite{Octavio_chevrel_2015} and it can be monovalent, divalent, trivalent or rare-earth elements. This leads to many family members and large space to tune their physical properties.
  Therefore, there have been lots of research  done on them, and many intriguing phenomena have been found and investigated.
  For example, SnMo$_6$S$_8$ and PbMo$_6$S$_8$ show superconductivity at low temperature with $T_c$ being about 14.2 and 15.2 K, respectively~\cite{Pb_superconductor_1973, Pb_Sn_superconductor_1985, Pb_Sn_Hc_1995}. 
  $\BaMoS$ and $\EuMoS$ have been found to be superconducting under pressure around several GPa~\cite{Ba_pressure_induced_sc_1982, BaEuSnMoS_HP_1988, Upper_Hc2_BaMoS_1989,EuMoS_SC_HP_1983,EuMoS_SC_HP_1984}.
  Ca$_x$Mo$_6$S$_8$ becomes superconducting when Ca vacancy is introduced with $x$ = 0.94~\cite{CaMoS_superconductivity}.
  Moreover, many of them have upper critical fields $H_{\rm c2}$ higher than 25 T and even up to 60 T~\cite{Pb_Sn_Hc_1995, Upper_Hc2_BaMoS_1989, EuMoS_SC_HP_1984}, which violates the Pauli paramagnetic limit in the weak coupling Bardeen-Schrieffer-Copper superconductor~\cite{NbRhC_2021,tinkham2004introduction}. 
  One of the possible reason is the strong spin-orbit coupling (SOC) of the electrons around the Fermi level, since Mo 4$d$ bands have been found to constitute the Fermi surface.
 In addition, we noticed that the resistance-temperature curve of $\BaMoS$ shows an anomalous behavior like ZrTe$_5$~\cite{MEUL1986,ROSSEL1985381,ZrTe5_R_1, ZrTe5_R_2}, which has sensitive topological phase transitions tuned by small external stimuli~\cite{ZrTe5_weng_2014,ZhouXingJiang_ZrTe5_2017,Light_ZrTe5_Dirac_2020,ZrTe5_light_weyl_Luo2021}.
 Thus, to understand these exotic physical properties, the extensive studies on their electronic structure and band topology are necessary and meaningful.
 
    \begin{figure*}
      \centering
      \includegraphics[width=0.8\linewidth]{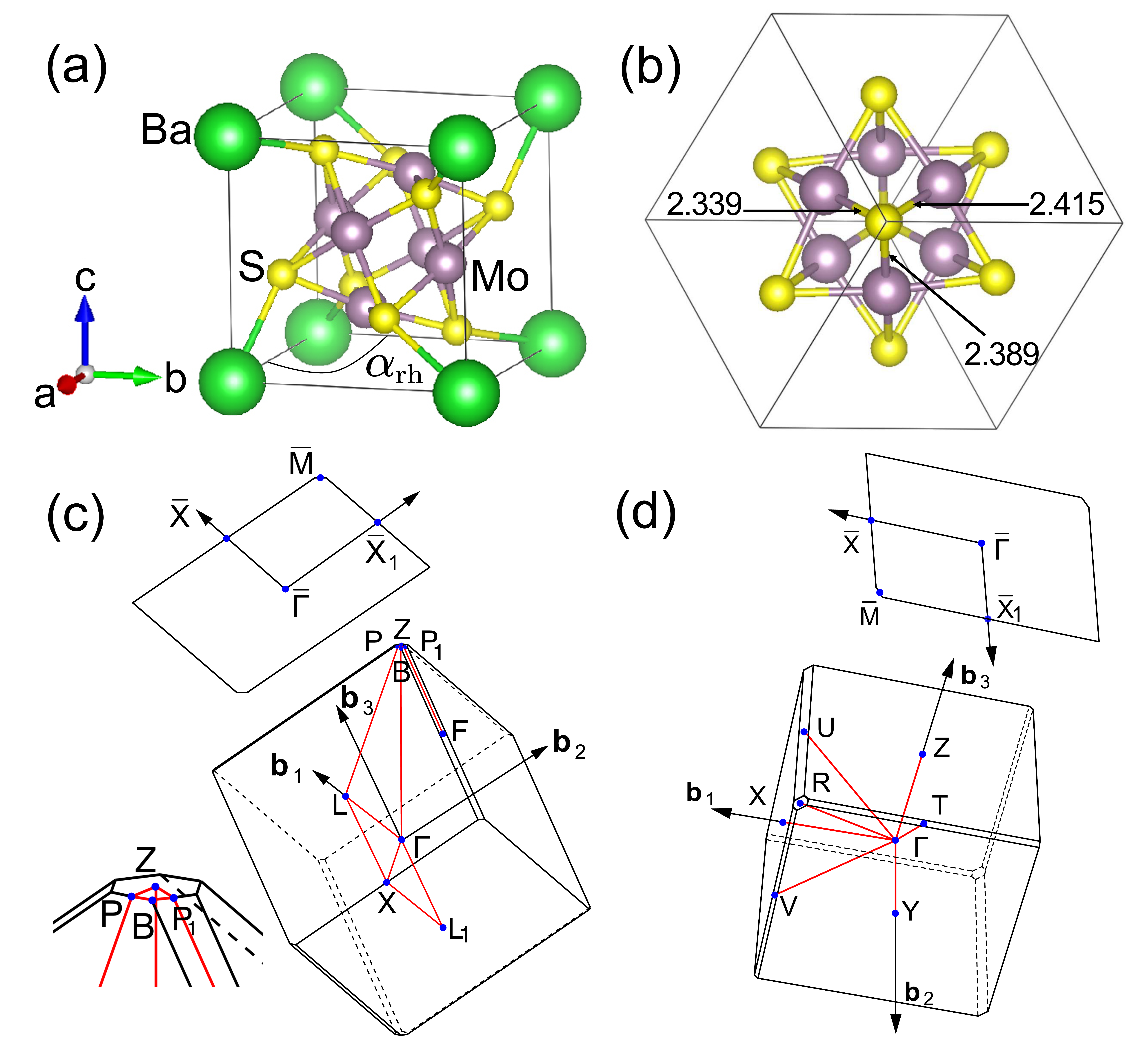}
      \caption{
         Crystal structures (primitive unit cell) and Brillouin zones of $\BaMoS$. 
         (a) The crystal structure of the $R\bar{3}$ phase, where the green balls represent Ba and the purple (yellow) represents Mo (S). 
         (b) The crystal structure of the $P\bar{1}$ phase in view of $\langle 111 \rangle$ direction with Ba atoms omitted for clearness. The numbers are three Mo-S bond lengths in unit of \AA.
         Panels (c) and (d) show the bulk and the $(001)$ surface Brillouin zones of the $R\bar{3}$ and the $P\bar{1}$ phases, respectively.
         The inset in Panel (c) magnifies the part around the Z point.
      }
      \label{fig:AM6S8_struct_bz}
    \end{figure*}
 
 In this work, we calculated and analyzed the topological states of some sulfides in these CP materials.
 We have chosen four representative compounds, namely Mo$_6$S$_8$ with $A$ site unoccupied, $\BaMoS$ and $\SrMoS$ with $A$ site occupied by nonmagnetic divalent ions, and $\EuMoS$ with $A$ site occupied by magnetic rare-earth ions Eu.
 We found that some of them have nontrivial band topology and further studied their dependence on structural and magnetic phase transitions, as well as different occupation cases.
 On considering their unconventional superconductivity, the existence of nontrivial band topology will be very appealing, and their mutual coupling may induce topological superconductivity and Majorana modes, such as those proposed in FeTe$_{1-x}$Se$_x$~\cite{Wang_FeTeSe_2015,Ding_FeTeSe_2018}, the HfRuP family~\cite{Qian2019}, and YCoC$_2$~\cite{YCoC2Xu}.

The rest of this paper is organized as follows:
First, we introduced the computational methods and software packages as well as the parameters we set. 
Second, we discussed the topological states of nonmagnetic compounds with $\BaMoS$ as the representative object.
Third, we studied the magnetic compound $\EuMoS$. Finally, we present a  discussion and our conclusions.

    \begin{figure*}
      \centering
      \includegraphics[width=0.9\linewidth]{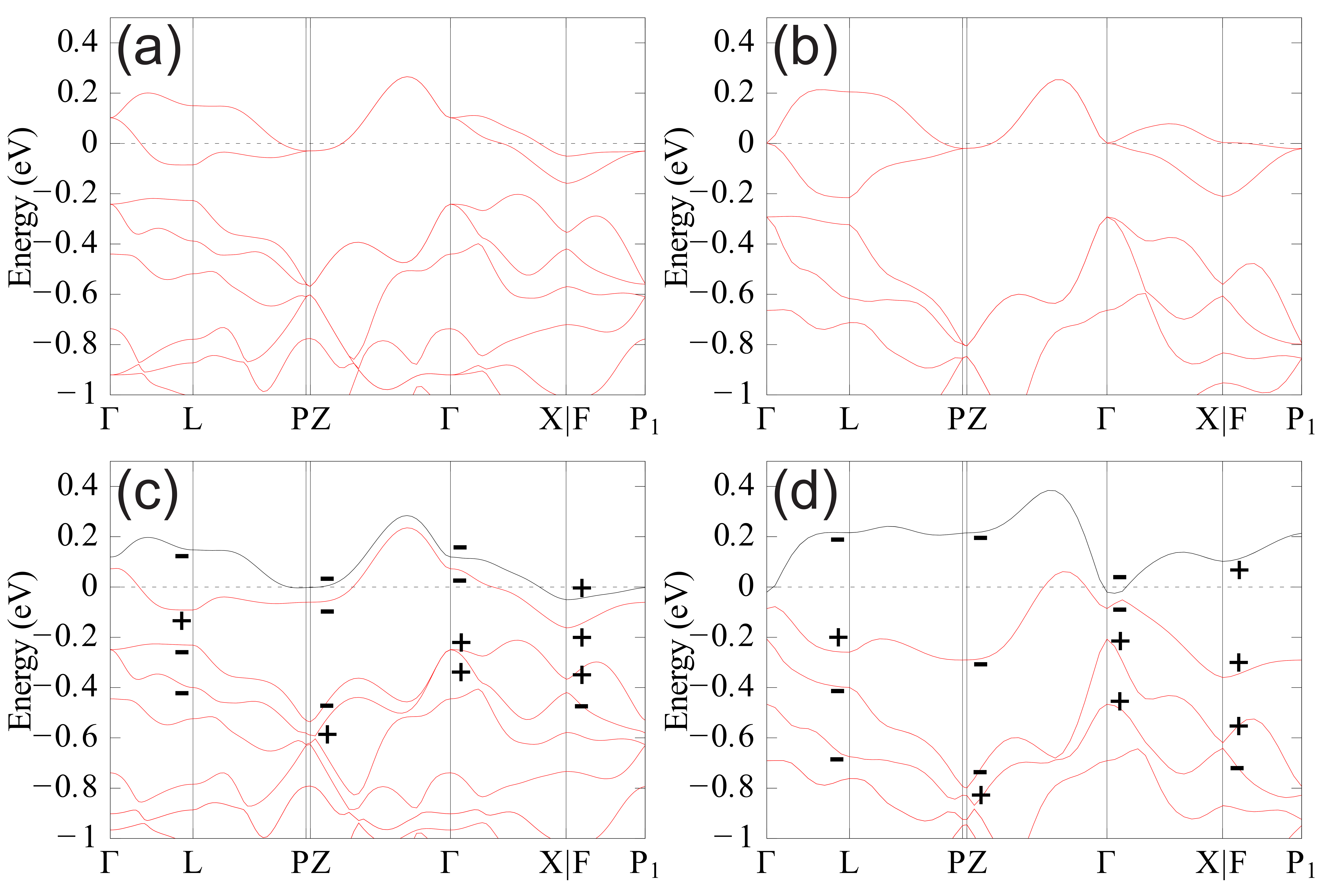}
      \caption{
        The band structures of $R\bar{3}$ phase $\BaMoS$ calculated with  non-SOC PBE (a),  non-SOC HSE06 (b), SOC PBE (c), and SOC HSE06 (d). 
        In Panels (c) and (d), ``-'' and ``+'' represent the parity eigenvalues of each Kramers degenerate pair at TRIMs and the red bands are the $N$ occupied ones, where $N$ represents the number of valence electrons of $\BaMoS$. Here, the coordinates of these TRIMs are $\Gamma$ (0, 0, 0), L (0.5, 0, 0), F (0.5, 0.5, 0), and Z (0.5, 0.5, 0.5).
      }
      \label{fig:BaMoS_R3_band}
    \end{figure*}

\section{Methodology}
  We employed the Vienna \textit{ab initio} simulation package  (\texttt{VASP})
~\cite{kresse1996efficient, KRESSE199615}
  to perform the electronic structure calculations. 
  The projector-augmented-wave (PAW) method~\cite{blochl1994projector,kresse1999ultrasoft} with the Perdew-Burke-Ernzerhof (PBE) 
  exchange-correlation functional~\cite{perdew_generalized_1996} was used.
  It is known that DFT calculation within local density approximation or generalized gradient approximation always underestimates the band gap and overestimates the band inversion. To amend this, hybrid functional HSE06~\cite{HSE06_1,HSE06_3} calculation is usually employed to check the band topology~\cite{Zunger_2011}.
  The plane-wave cutoff for kinetic energy was set as 500 eV, and a $7 \times 7 \times 7$ $\Gamma$-centered Monkhorst-Pack $k$-point mesh was used to sample the Brillouin zone (BZ) in the self-consistent charge convergence calculation.
  To investigate the topological states, we constructed the Wannier functional based effective Hamiltonian using the \texttt{WANNIER90} package~\cite{wannier90v3}, and performed the calculations of surface states and flow spectra of Wannier charge centers (WCCs) of occupied bands within Wilson loop scheme~\cite{Dai_Wilson_2011} using the \texttt{WANNIERTOOLS} package~\cite{WU2017}. This package is also used when we check the existence of the in-gap node points. We constructed the Wannier functions for the bands composed of Mo $d$ and S $p$ orbitals.

\section{Result and Discussion}

\subsection{$\BaMoS$}

The crystal structures and the BZs of $\BaMoS$ are shown in 
Fig.~\ref{fig:AM6S8_struct_bz}. As most of the CP materials, it is  $R\bar{3}$ phase at room temperatures and takes a structural phase transition to $P\bar{1}$ phase around 175 K~\cite{BaR3_P1_structure_1986} when temperature drops down. 
$R\bar{3}$ phase can be assumed as a slightly distorted cubic structure, in which the Mo$_6$S$_8$ cluster is centering the cubic lattice formed by Ba atoms.
The 90$\degree$ angle  between the  lattice vectors   is reduced to $88.711\degree$ ($\alpha_{\rm rh}$) to form the $R\bar{3}$ phase ~\cite{BaR3_P1_structure_1986}, as shown in Fig.~\ref{fig:AM6S8_struct_bz}(a).
The point group symmetry of $R\bar{3}$ phase is $C_{3i}$, where the $C_3$ rotation axis is along $\langle 111 \rangle$ direction in the rhombohedral lattice. As temperature drops down, the $C_{3}$ rotation symmetry is broken, and the symmetry reduces to $P\bar{1}$. This structural phase transition can be seen in the changes of the Mo-S bond lengths indicated in Fig.~\ref{fig:AM6S8_struct_bz}(b), where the three Mo-S bond lengths are different.
The size of ions at $A$ sites will affect the amplitude of distortion
from $C_{3}$ symmetry.
This distortion of $\MoS$ cluster is a kind of cooperative Jahn-Teller distortion~\cite{GaTa4Se8_2020} to form $P\bar{1}$ phase.
All the crystal structures~\cite{BaR3_P1_structure_1986,Sr_R3_structure,Sr_P1_structure,MoS_springer,MoS_structure,MoS2_struct,Eu_R3_structure_3,Eu_P1_structure} used in work are shown in the Supplementary Material~\cite{BaMoS_supp}.

    \begin{figure}
      \centering
      \includegraphics[width=1.0\linewidth]{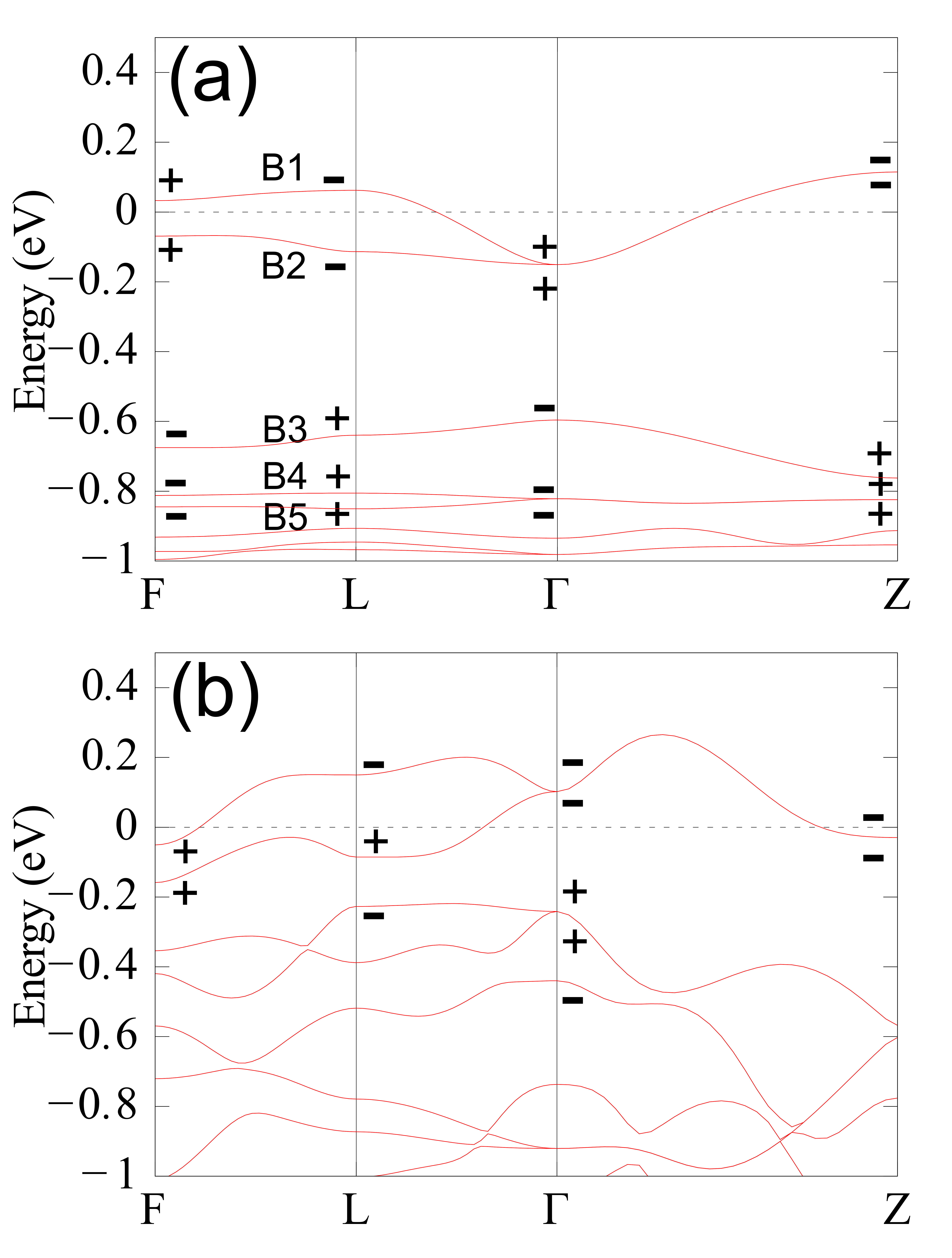}
      \caption{
        (a) The non-SOC band structure calculated with scaling the lattice constants of $R\bar{3}$ phase $\BaMoS$ by 128 \% and fixing the Mo$_6$S$_8$ as a rigid cluster. ``B1'' ``B2'' and so on label each flat band. The parity eigenvalues at TRIMs are also labeled.
        (b) The non-SOC band structure of the  real  $\BaMoS$.
      } 
      \label{fig:BaMoS_inversion_band}
    \end{figure}

    \begin{figure*}
      \centering
      \includegraphics[width=1.0\linewidth]{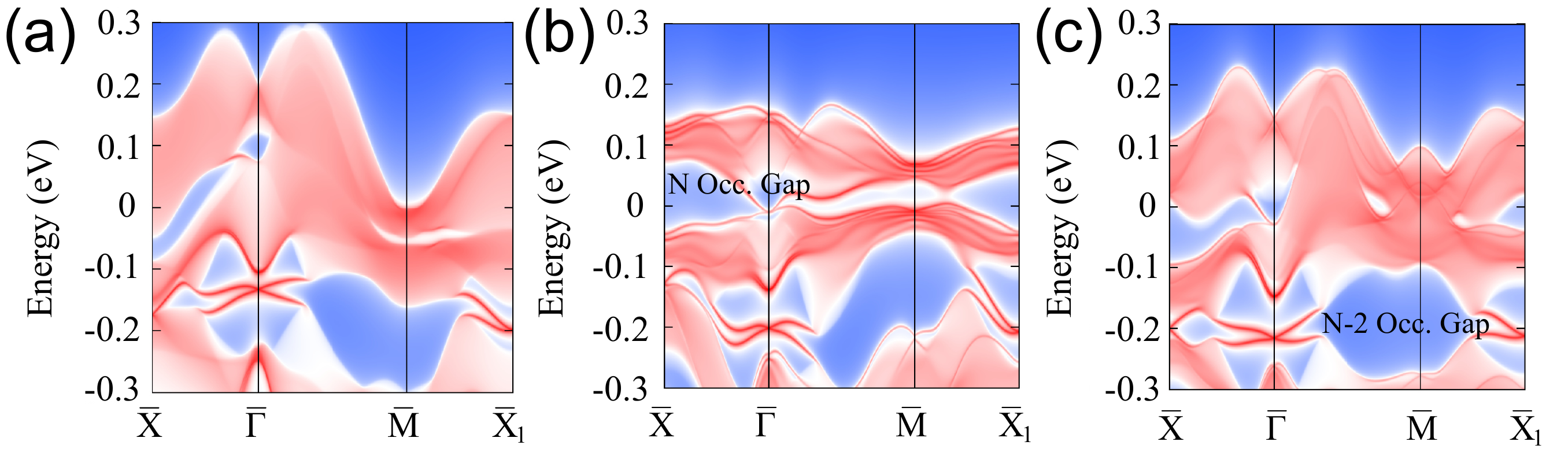}
      \caption{
           The $(001)$ surface states of $R\bar{3}$ phase $\BaMoS$ with SOC. 
           (a) The surface states calculated with the original band structure.
           (b) The surface states calculated with the renormalized band structure, which has a global gap in the $N$ occupation case.
           (c) The surface states calculated with the renormalized band structure, which has a global gap in the $N-2$ occupation case.
      }
      \label{fig:BaMoS_surface_148_norm}
    \end{figure*}
    
    \begin{figure}
      \centering
      \includegraphics[width=1.0\linewidth]{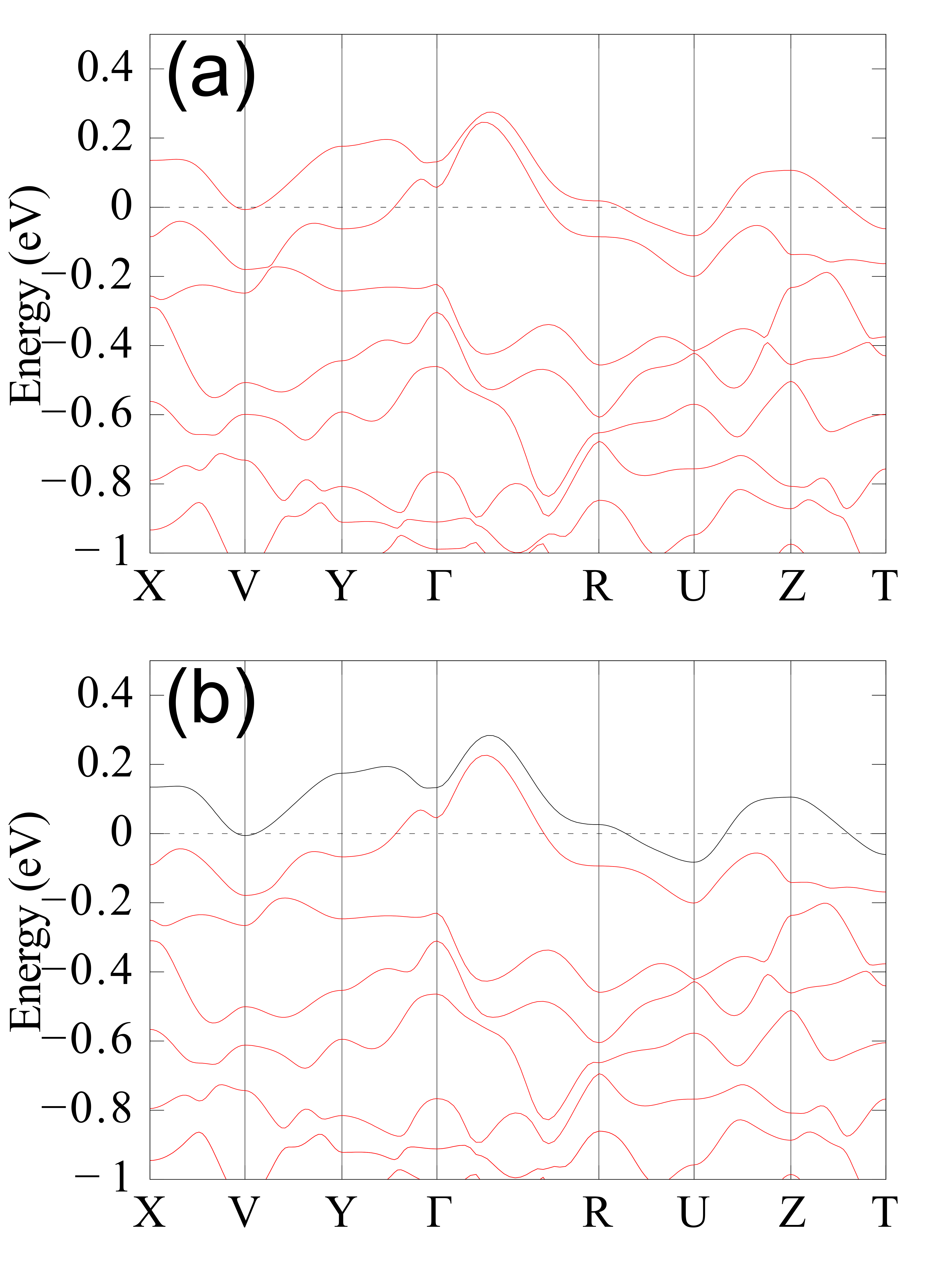}
      \caption{
        The band structures of $P\bar{1}$ phase $\BaMoS$ without SOC (a) and with SOC (b).
        In panel (b), the red bands are the $N$ occupied bands, where $N$ represents the  number of valence electrons of $\BaMoS$.
      }
      \label{fig:BaMoS_P1_band}
    \end{figure}

Now, let us consider the electronic structures and the topological states of the $R\bar{3}$ phase.
In Fig.~\ref{fig:BaMoS_R3_band}, the bands around the Fermi level are mainly from Mo 4$d$ orbitals and they have been proposed to be understood in the basis set composed by molecular orbitals of Mo$_6$ octahedral cluster~\cite{nohl1982band}.
In the non-SOC case, as shown in Fig.~\ref{fig:BaMoS_R3_band}(a), the two bands  crossing the Fermi level are two-fold degenerated along the $\Gamma-$Z path, 
which is  protected by the $C_3$ rotation symmetry, and their representation is $^1E\oplus^2E$.
These two irreducible representations $^1E$ and $^2E$ are complex conjugate to each other. By taking the time-reversal symmetry into account, they will combine together to form a two dimensional irreducible representation of the type II magnetic group~\cite{el-batanouny_wooten_2008}.
After including SOC, as shown in Fig.~\ref{fig:BaMoS_R3_band}(c), the splitting in these two bands is obvious, namely there are finite band gaps between these two bands (each has Kramers degeneracy) at any $k$ point in the BZ.
For this reason,  SOC  is necessary for getting the possible band insulator state in $R\bar{3}$ phase $\BaMoS$. Otherwise, it would be a filling-enforced metal~\cite{Watanabe_filling_metal_2016,Chen2018}.

Now let us consider the topological states with SOC.
For convenience, we denote $N$ to represent the number of valence electrons in the case of $A$ being a divalent ion with $x=1$, such as $\BaMoS$. There are $N-2$ valence electrons when $A$ site is empty. 
When SOC is considered, the Kramers degenerated $(N-1)$- and $N$-th bands are gapped from the $(N+1)$- and $(N+2)$-th bands at each $k$-point as we mentioned before.
According to the Fu-Kane formula~\cite{FuKane_2007}, we can get the topological  $\mathbb{Z}_2$ indices $(v_0;v_1,v_2,v_3)$ from the parity eigenvalues of all occupied Kramers degenerate pairs on time-reversal invariant momenta (TRIMs). 

We found if all the lowest $N+2$ bands were occupied, the $\mathbb{Z}_2$ indices are $(0;0,0,0)$.
If all the lowest $N$ bands are occupied (i.e., the $N$ occupation case), the $\mathbb{Z}_2$ indices are $(1;0,0,0)$, which means the $\mathbb{Z}_2$ indices for Kramers degenerated  $(N+1)$- and $(N+2)$-th bands are $(1;0,0,0)$.
If all the lowest $N-2$ bands are occupied (i.e., the $N-2$ occupation case, namely $A$ is unoccupied), the $\mathbb{Z}_2$ indices are $(1;1,1,1)$, which means the  $\mathbb{Z}_2$ indices for the  $(N-1)$- and $(N)$-th bands are $(0;1,1,1)$.
From these $\mathbb{Z}_2$ indices, 
the gap between the group of the lowest $N-2$ bands and that of  $N-1$ to $N+2$ bands
is topologically non-trivial. That means there is band inversion happening between these two groups, which indicates there must be nodal line(s) in this gap in the non-SOC case~\cite{Song_nodalline}.
The detailed calculations including the parity distribution and WCCs are shown in the Supplementary Material~\cite{BaMoS_supp}.

The band structures with HSE06 functional are shown in Figs.~\ref{fig:BaMoS_R3_band}(b) and ~\ref{fig:BaMoS_R3_band}(d). In the non-SOC + HSE06 case, the band structure is much more like a semimetal than the PBE case.
But the degeneracy of the bands crossing the Fermi level along the $\Gamma-$Z path remains. 
In the SOC + HSE06 case, such degeneracy is broken, and the band gap between them is much more enlarged than PBE case with small electron and hole pockets left. Comparing the parities in  both cases, the parity eigenvalues and their distribution of the $N$-th and the ($N+2$)-th bands are not changed as shown in Figs.~\ref{fig:BaMoS_R3_band}(c) and~\ref{fig:BaMoS_R3_band}(d). So the topological states from HSE06 functional are not changed.

For a better understanding of the topological states in both the $N-2$ occupation and the $N$ occupation cases, we investigated the band inversion mechanism, which provides an intuitive picture of topological phase transitions.
Let us consider the non-SOC condition first. By scaling the lattice vectors by 128 \% and fixing the $\MoS$ cluster as a rigid molecular centering on the (0.5, 0.5, 0.5) inversion center, the hopping between the clusters have been reduced and we got topologically trivial flat bands as shown in Fig.~\ref{fig:BaMoS_inversion_band}(a). We labeled the flat bands as ``B1,'' ``B2,'' and so on. 
B1 and B2 can be represented as the elementary band representation (eBR)  $^1E_g ^2E_g @ 3b$.
B3 is from eBR $A_u @ 3b$, and
B4 and B5 are from eBR $^1E_u ^2E_u @ 3b$.

To clarify  the topological states in both  the $N$ occupation and the $N-2$ occupation cases, we only need to tell  which bands (i.e., B3, B4, and B5) inverted with the B1 and B2 in the real crystal because the other lower bands are not entangled with B1 and B2 until the lattice constants reduce to the real values. Comparing the parity eigenvalues at TRIMs (i.e., $\Gamma$, Z,  L and F points) in the flat bands and those in the real crystal, we can find the following: 

1) At the $\Gamma$ point,  B1, B2 (with ``+'' parity eigenvalues) and B4, B5 (with ``-'' parity eigenvalues) are inverted;

2) At the L point, the parities of B1 and B2 change from two negative values to one negative and one positive, which means there is band inversion happens. 
Comparing Figs.~\ref{fig:BaMoS_R3_band}(c) and~\ref{fig:BaMoS_inversion_band}(b), we can find that SOC can only lift the band degeneracy but not affect the parity distribution at TRIMs. 
The hopping between the molecular orbitals from $\MoS$ clusters is crucial to the band inversion.

We further calculated the $(001)$ (rhombohedral basis) surface states of $R\bar{3}$ phase $\BaMoS$ with SOC as shown in Fig.~\ref{fig:BaMoS_surface_148_norm}(a).
Because of the absence of global gap in both $N$ and $N-2$ occupation cases, the surface Dirac cone is mixed with the bulk states. 
Therefore, we artificially shifted all the \textit{ab initio} bands at every $k$ point by a $k$-dependent energy value, which does not change the energy order of the bands and the band topology,
but it results in a global band gap.
The details of this method are described in the Supplementary Material~\cite{BaMoS_supp}. For later convenience, we call this procedure  ``renormalization''.
We constructed two renormalized band structures which have global gap in the $N$ and $N-2$ occupation cases, respectively. 
Based on these renormalized band structures, we calculated the corresponding surface states with 0.02 and 0.01 eV onsite energy correction to the atoms in the surface region for the $N$ occupation and $N-2$ occupation cases, respectively. 
As shown in Figs.~\ref{fig:BaMoS_surface_148_norm}(b) and ~\ref{fig:BaMoS_surface_148_norm}(c), there is a global gap in bulk states and clear surface Dirac cones in both cases. 
In the $N$ occupation case, there is only one surface Dirac cone at $\bar{\Gamma}$ and there are three surface Dirac cones in $N-2$ occupation case at $\bar{\rm X}$, $\bar{\Gamma}$ and $\bar{\rm X}_1$. 
The number and the distribution of surface Dirac cones are consistent with the parity distribution.

    \begin{figure*}
      \centering
      \includegraphics[width=1.0\linewidth]{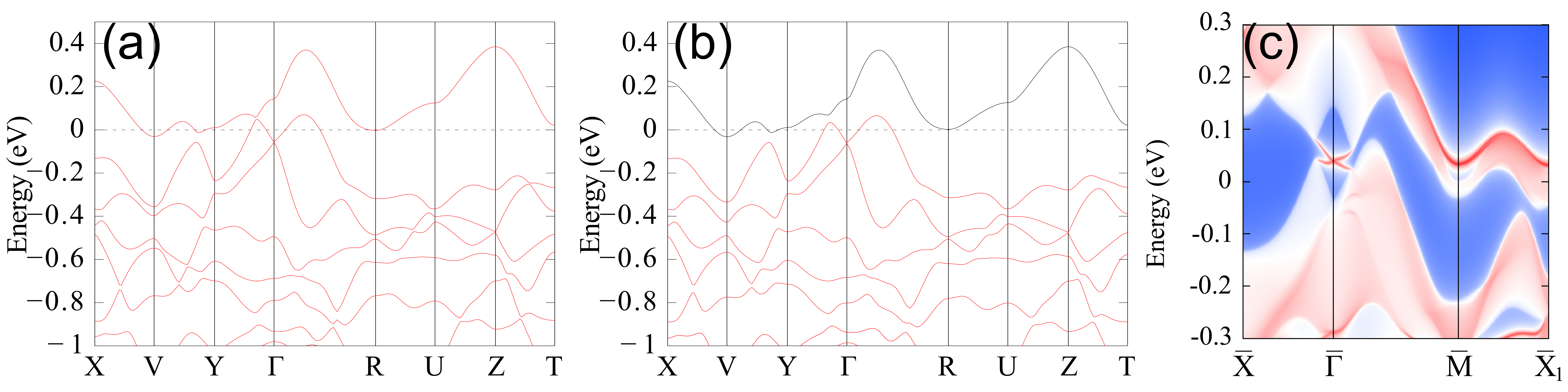}
      \caption{
        The band structures without SOC (a) and with SOC (b) of $P\bar{1}$ phase $\SrMoS$. 
        In panel (b), the red bands are the $N$ occupied ones, where $N$ represents the number of the valence electrons of $\BaMoS$.
        (c) The $(001)$ surface states of $P\bar{1}$ phase $\SrMoS$ with SOC.
      }
      \label{fig:SrMoS_band_2}
    \end{figure*}

Now let us consider $P\bar{1}$ phase $\BaMoS$. 
The electronic structure of $P\bar{1}$ is quite similar to the $R\bar{3}$ phase. The obvious difference is that there is no double degenerated band in the non-SOC case due to the breaking of $C_3$ rotation symmetry, as shown in Fig.~\ref{fig:BaMoS_P1_band}(a). With the inversion symmetry retained, we can still get their topological states from Fu-Kane formula.
Due to the tiny difference in both the crystal and electronic structures between $R\bar{3}$ and $P\bar{1}$ phases, the topological $\mathbb{Z}_2$ indices remain unchanged in both of the $N-2$ occupation and $N$ occupation cases. The detailed information (including the parity distribution, WCCs, and surface states) is shown in the Supplementary Material~\cite{BaMoS_supp}.

\subsection{$\SrMoS$ and $\MoS$}

In the $R\bar{3}$ phase, the crystal structure and the band structure of $\SrMoS$ are basically the same as $R\bar{3}$ phase $\BaMoS$, except for the larger bandwidth of $\SrMoS$ due to the smaller lattice constants.
The topological $\mathbb{Z}_2$ indices of $R\bar{3}$ phase $\SrMoS$ are the same as $R\bar{3}$ phase $\BaMoS$ in both of the $N-2$ occupation and the $N$ occupation cases.
We do not repeat the discussion in the main text. The band structures and the information of the topological states of $R\bar{3}$ phase $\SrMoS$ are shown in the Supplementary Material~\cite{BaMoS_supp}.

As shown in Figs.~\ref{fig:SrMoS_band_2}(a) and~\ref{fig:SrMoS_band_2}(b) for $P\bar{1}$ phase $\SrMoS$, the gap between the $N$- and the $(N+2)$-th bands at $\Gamma$ point
is much larger than the one in $P\bar{1}$ phase $\BaMoS$, and there are only small electron- and hole- pockets near the Fermi level.
This may be attributed to the subtle difference between the structures of $\BaMoS$ and $\SrMoS$ in the $P\bar{1}$ phase, as shown in the Supplementary Material~\cite{BaMoS_supp}.

In the $N$ occupation case, the topological $\mathbb{Z}_2$ indices with SOC are $(1;0,0,0)$, which are the same as $P\bar{1}$ phase $\BaMoS$.
The larger bulk band gap at $\Gamma$ leads to a clear surface Dirac cone as shown in Fig.~\ref{fig:SrMoS_band_2}(c), which is calculated normally without above renormalization procedure.
But for the $N-2$ occupation case, the topological $\mathbb{Z}_2$ indices with SOC are $(0;1,1,1)$.
They are different from those of $P\bar{1}$ phase $\BaMoS$.
The nearly closing  band gap at the $\Gamma$ point indicates $\SrMoS$ is close to the topological phase transition point from $(1;1,1,1)$ to $(0;1,1,1)$ due to the  crystal distortion induced by replacing Ba with smaller Sr ions.

    \begin{figure}
      \centering
      \includegraphics[width=0.9\linewidth]{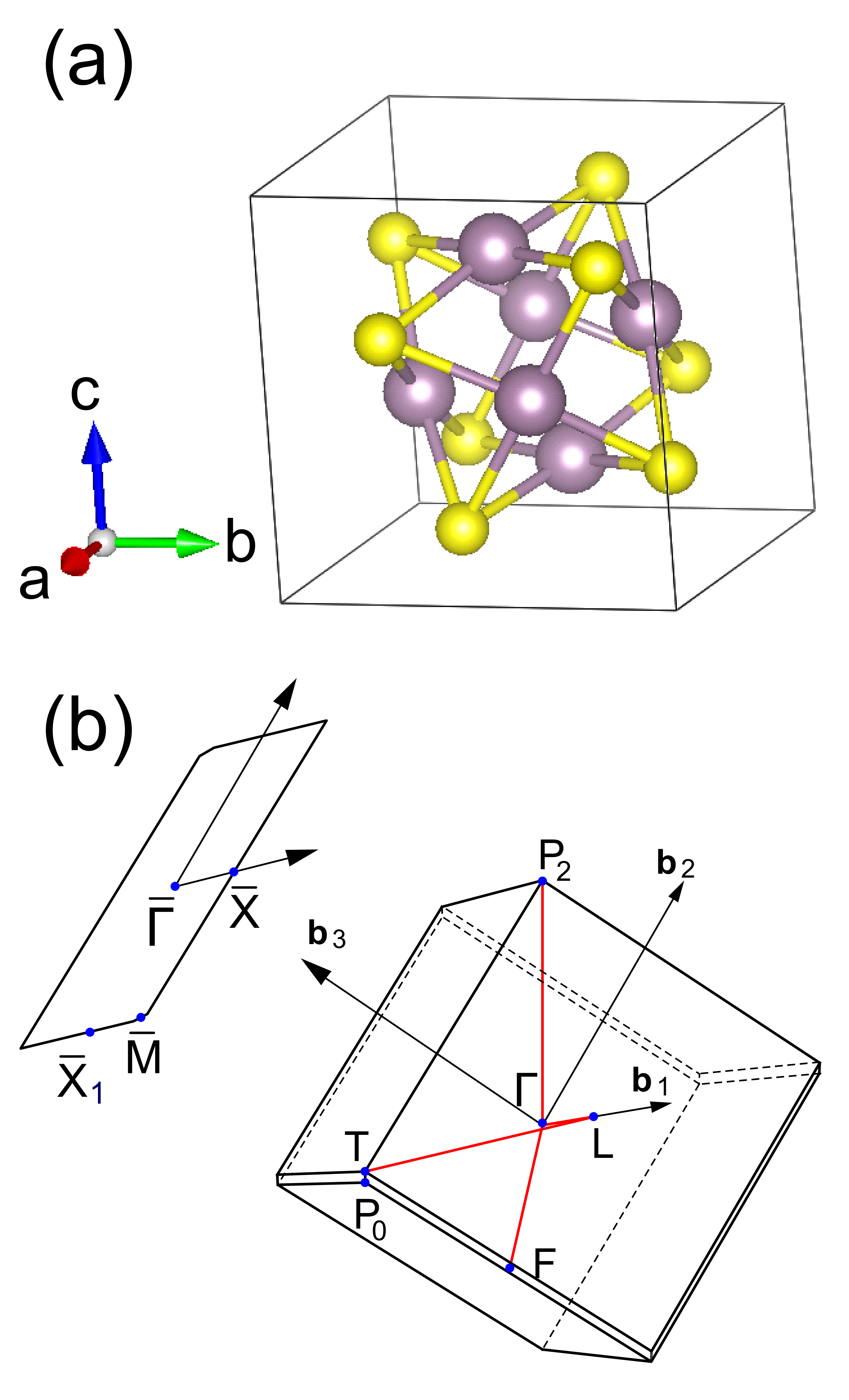}
      \caption{
         Crystal structure (a) and Brillouin zone (b) of $R\bar{3}$ phase Mo$_6$S$_8$. The purple atoms and yellow atoms represent Mo and S, respectively.
      }
      \label{fig:MoS_struct_bz}
    \end{figure}

Because the topological state of $\BaMoS$ in the $N-2$ occupation case is a strong topological insulator, it is natural to reduce two valence electrons by removing the divalent metal atoms (i.e., Ba) to get a topological insulator with a large gap near the Fermi level. This is exactly $\MoS$ with the $A$ site empty.
In the $R\bar{3}$ phase, $\MoS$ has a similar crystal structure to that of $\BaMoS$. The lattice constant is 6.428 \AA \ and  $\alpha_{\text{rh}}$ is large then $90\degree$ (i.e., 91.250$\degree$)~\cite{MoS_springer,MoS_structure}. 
Thus, the Brillouin zone is different from  $\BaMoS$ which have divalent metal atoms. The crystal structure and the Brillouin zone of $R\bar{3}$ phase $\MoS$ are shown in Fig.~\ref{fig:MoS_struct_bz}. 

\begin{figure}
      \centering
      \includegraphics[width=1.0\linewidth]{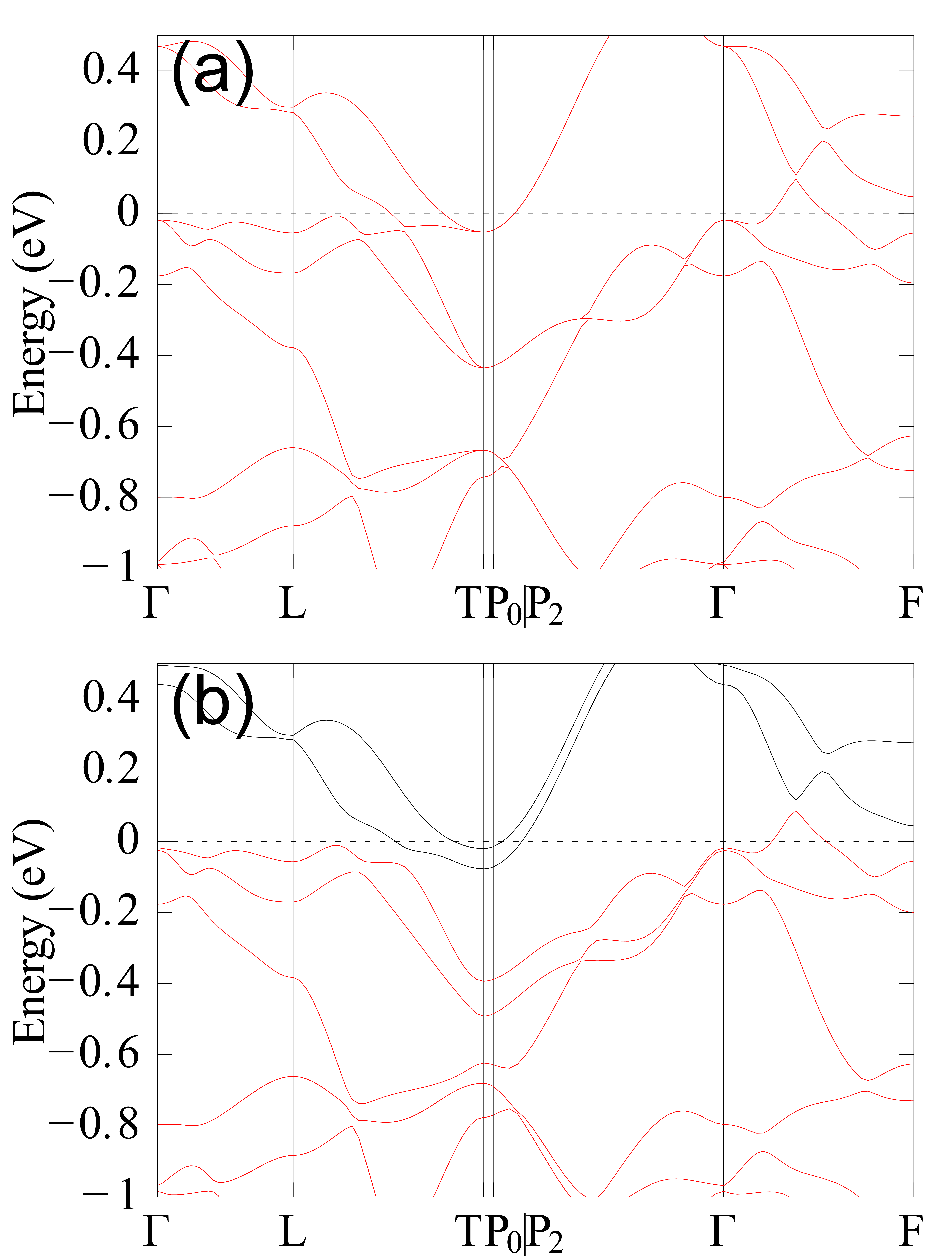}
      \caption{
        The band structures of $R\bar{3}$ phase Mo$_6$S$_8$ without SOC (a) and with SOC (b).
        In panel (b), the red bands are the $N-2$ occupied ones, where $N$ represents the number of valence electrons of $\BaMoS$.}
      \label{fig:MoS_band_148}
    \end{figure}

But from the band structures shown in Fig. \ref{fig:MoS_band_148}, the gap is not as large as we expected before. 
The $\mathbb{Z}_2$ indices with SOC are $(1;1,1,1)$, which are the same as those of $\BaMoS$ in the $N-2$ occupation case. The detailed information can be found in the Supplementary Material~\cite{BaMoS_supp}.

\begin{figure}
      \centering
      \includegraphics[width=1.0\linewidth]{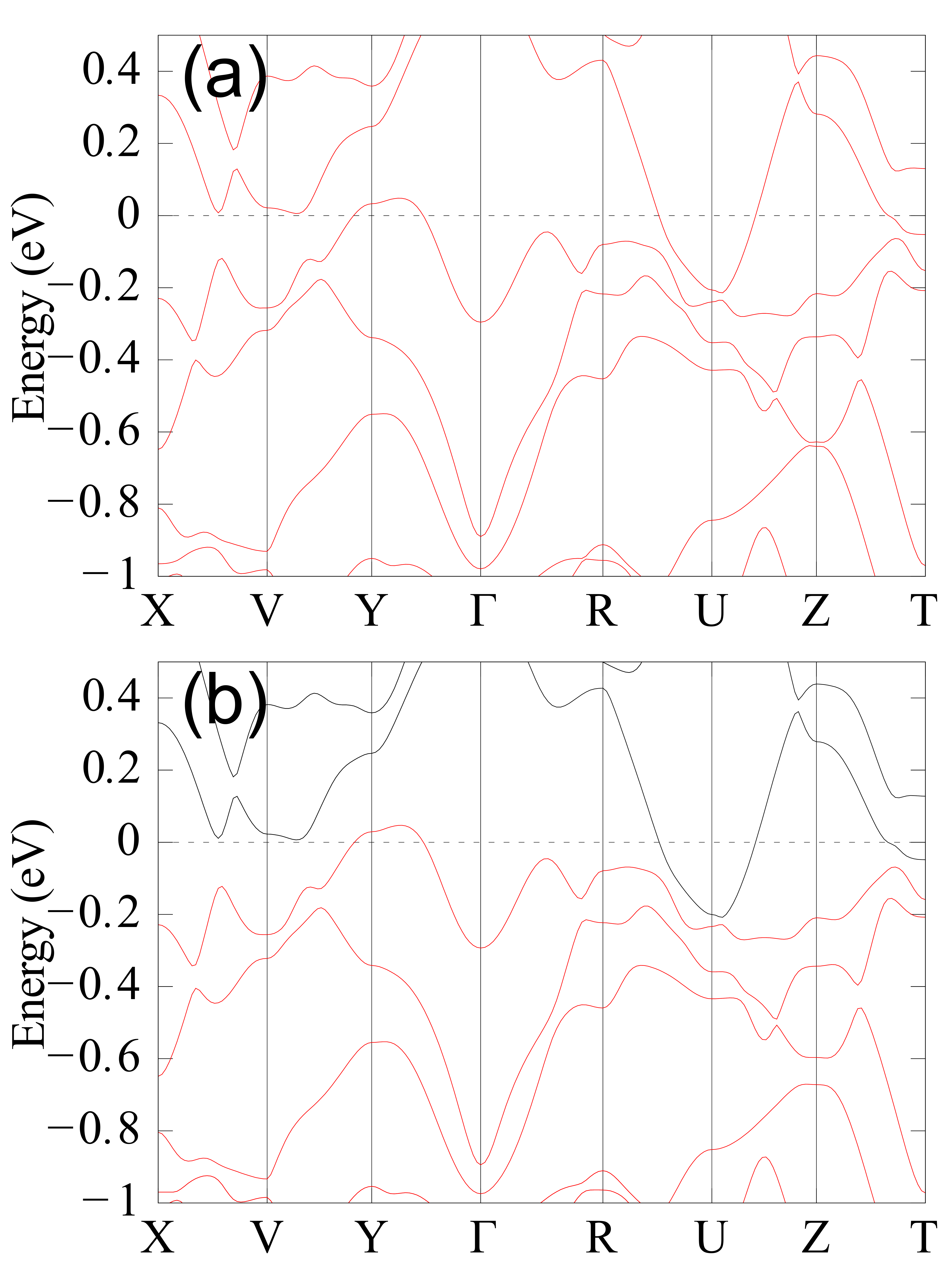}
      \caption{
      The band structures of $P\bar{1}$ phase $\MoS$ without SOC (a) and with SOC (b).
      In panel (b), the red bands are the $N-2$ occupied ones, where $N$ represents the number of valence electrons of $\BaMoS$.
      }
      \label{fig:MoS_2_bands}
    \end{figure}

For the $P\bar{1}$ phase,  $\MoS$ has quite different crystal structure from $\BaMoS$ and $\SrMoS$ and the electronic structure is also quite different from them, as shown in Fig.~\ref{fig:MoS_2_bands}.
The $\mathbb{Z}_2$ indices with SOC are $(1;0,0,1)$, which are different from $\BaMoS$ and $\SrMoS$ in the $P\bar{1}$ phase and the $N-2$ occupation case.
The parity distribution is shown in the Supplementary Material~\cite{BaMoS_supp}.

\subsection{$\EuMoS$}
$\EuMoS$ has magnetic order at low temperatures (less than 0.4 K), but it seems not to be the simple ferromagnetic order~\cite{EuMoS_magnetism_1984}. 
The seven 4$f$ electrons on Eu$^{2+}$ ions are fully spin polarized and completely localized. In this sense, EuMo6S8 and BaMo6S8 are very similar, except for the breaking of the time-reversal symmetry and the Kramers degeneracy.
Thus, we expect that the $\mathbb{Z}_2$ topological insulator in $\BaMoS$ can be driven to axion insulator~\cite{Axion_1977, PhysRevLett.102.146805_2009, PhysRevLett.119.246401_2017, Axion_2009, EuIn2As2_2019}, Weyl semimetal~\cite{TaAs_hmweng2015,TaAs_exp,Wan_mag_weyl,HgCrSe_weyl2, 2021arXiv211204127H} or trivial magnetic insulator in $\EuMoS$.
We assumed an artificial ferromagnetic order for both the $R\bar{3}$  and $P\bar{1}$ phases to investigate their topological states.

\begin{figure}
      \centering
      \includegraphics[width=1.0\linewidth]{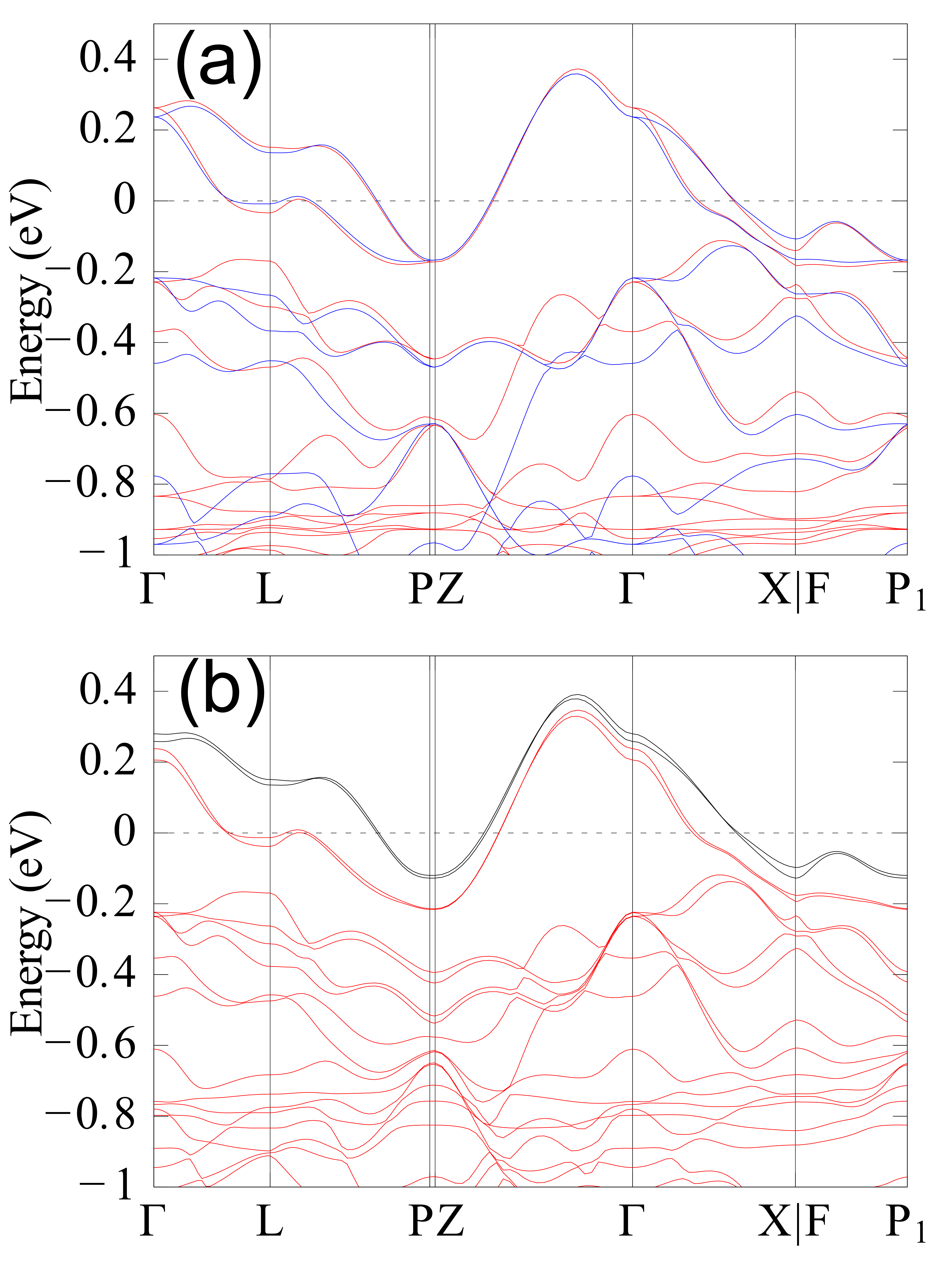}
      \caption{
        The band structures of $R\bar{3}$ phase EuMo$_6$S$_8$  without SOC (a) and with SOC (b) in ferromagnetic order.
        In panel (a), the red  blue colors represent the spin up and spin down, respectively.
        In panel (b), the red bands are the $N$ occupied ones, where $N$ represents the number of valence electrons of $\BaMoS$. 
        Here, the seven 4$f$ electrons of Eu$^{2+}$ are localized and not counted into the valence electrons.
      }
      \label{fig:EuMoS_band_148}
    \end{figure}

Now let us consider the $R\bar{3}$ phase firstly.
Figure ~\ref{fig:EuMoS_band_148} shows the LDA+U band structures of EuMo$_6$S$_8$ in the ferromagnetic order.
The on-site Hubbard $U = 5$ eV on the $f$ orbitals of Eu is considered. 
The difference between $\EuMoS$ and $\BaMoS$ is the spin splitting in the band structure. The spin splitting is quite small and it does not introduce any band inversion. Therefore, the topological states of $\EuMoS$ in both the $N$ and $N-2$ occupation cases are axion insulators as indicated by $z_4=2$,
which is calculated with SOC according to the formula~\cite{Ashvin_2012, MTQC_2018, watanabe_MSBI_2018,  Xu2020}
\begin{equation}
z_{4}=\sum_{K} n^{-}_{K}\ {\rm mod} \ 4,
\end{equation}
where $n^{-}_{K}$ is the number of occupied bands with odd parity eigenvalues at inversion-invariant momentum $K$.
We also  checked our results with increasing $U$ to 6, 7 and 8 eV, and they are not changed.


\begin{figure}
      \centering
      \includegraphics[width=1.0\linewidth]{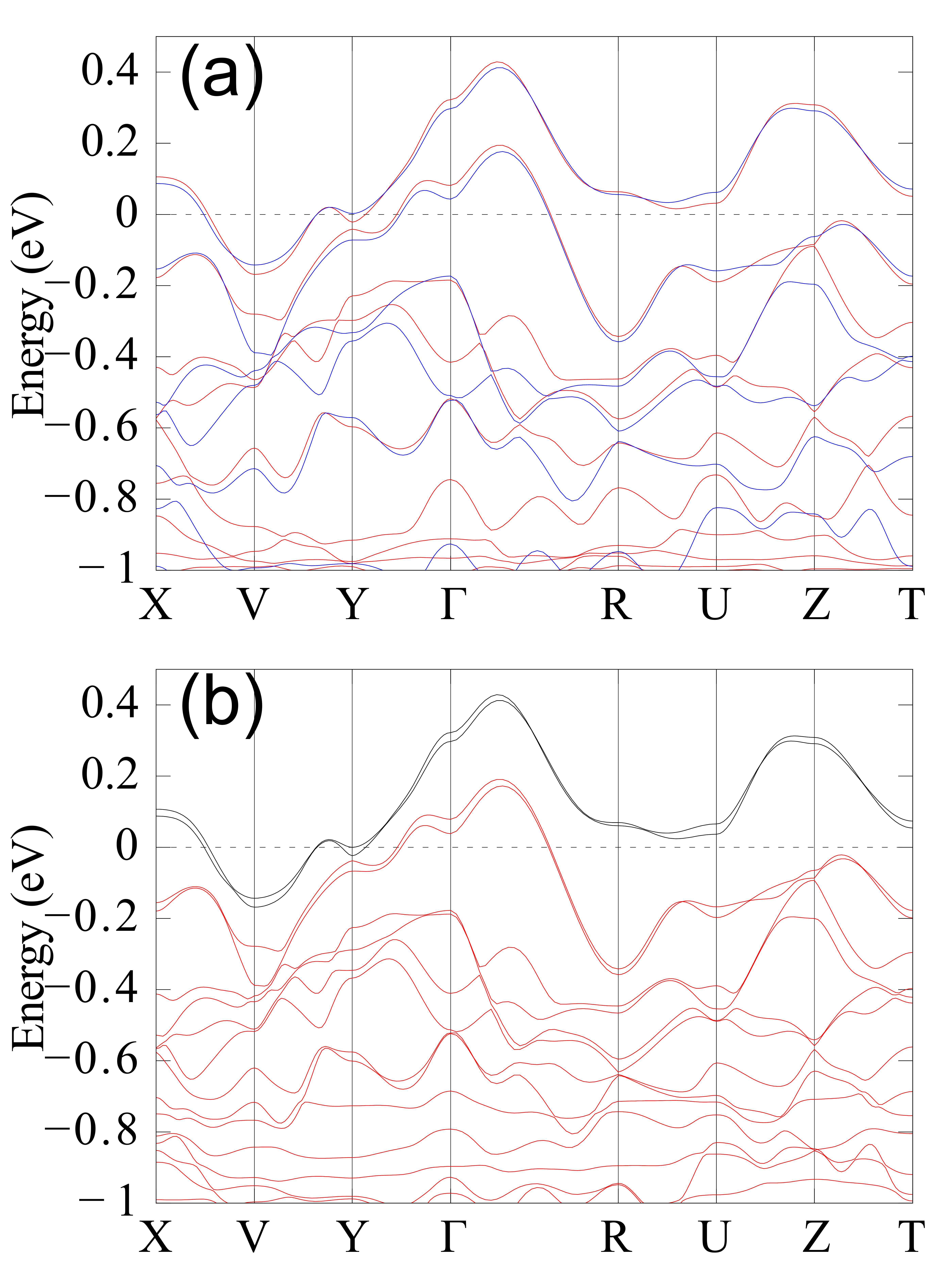}
      \caption{
        The band structures of $P\bar{1}$ phase EuMo$_6$S$_8$ without SOC (a) and with SOC (b) in ferromagnetic order.
        In panel (a), the red and blue colors represent the spin up and spin down, respectively.
        In panel (b), the red bands are the $N$ occupied ones, where $N$ represents the number of valence electrons of $\BaMoS$.
        Here, the seven 4$f$ electrons of Eu$^{2+}$ are localized and not counted into the valence electrons.
      }
      \label{fig:EuMoS_band_2}
    \end{figure}

For the $P\bar{1}$ phase, the band structures are shown in Figs.~\ref{fig:EuMoS_band_2}(a) and \ref{fig:EuMoS_band_2}(b).
In the $P\bar{1}$ phase, the $z_4$ indices with SOC are 0 in both of the $N-2$ and $N$ occupation cases.
This difference comes from the additional band inversion caused by structural phase transition from $R\bar{3}$ to $P\bar{1}$ phase.
For the $N$ occupation case,  at Y (0.0, 0.5, 0.0), the $(N-1)$- and $N$-th bands are inverted with the $(N+1)$- and $(N+2)$-th bands referring to the $R\bar{3}$ case. The $(N-1)$- and $N$-th bands and the $(N+1)$- and $(N+2)$-th bands have different parity eigenvalues, leading to the topological trivial state.
For the $N-2$ occupation case, the number of occupied bands with odd parities is reduced by two at V (0.5, 0.5, 0.0), 
and this results in the topological trivial state.
Similarly, the topological state of $P\bar{1}$ phase is not changed with increasing $U$	 to 6, 7 and 8 eV.

\section{Conclusion and discussion}
We systematically studied the topological states of CP materials.
For $\BaMoS$ in both of the $R\bar{3}$ and $P\bar{1}$ phases, the topological $\mathbb{Z}_2$ indices are $(1;0,0,0)$ in the $N$ occupation and $(1;1,1,1)$ in the $N-2$ occupation cases. 
For $R\bar{3}$ phase $\SrMoS$, the topological states are the same as $R\bar{3}$ phase $\BaMoS$.
For $P\bar{1}$ phase $\SrMoS$, it is a strong topological insulator in the $N$ occupation case, with $\mathbb{Z}_2=(1;0,0,0)$. For the $N-2$ occupation case, it is close to a topological phase transition point between the strong and weak topological states.
For $\MoS$ without the divalent metal atoms, it is a topological insulator and the topological invariants are the same as 
$R\bar{3}$ phase $\BaMoS$ in the $N-2$ occupation case.
But in the $P\bar{1}$ phase, the topological indices are $(1;0,0,1)$, which are different from the previous compounds because of the further crystal distortion without cation ions $A$. 
For $\EuMoS$ with ferromagnetic order, it is an axion insulator with $z_4=2$ in both  the $N$ and $N-2$ occupation cases in the $R\bar{3}$ phase, whereas it is a trivial magnetic insulator in the $P\bar{1}$ phase because of the additional band inversion.
The information of topological states of the compounds with time-reversal symmetry
can also be seen in the results of the high-throughput searching of topological materials~\cite{TQC_2017,Bernevig_Topodata_2019,Bernevig_Topodata_web_2019,Zhang_Topodata_2019,Wan_Topodata_2019}. However, some of the results are different from ours because of the difference in detail of calculations, such as the crystal structures and the exchange-correlation functionals.
Combining the intriguing physical properties of CP materials, such as high transition temperature and high upper critical field superconductivity,
the nontrivial band topology revealed in this work may open another dimension in their research and provide a platform to construct topological superconductivity and Majorana modes.

\begin{acknowledgments}
We acknowledge the supports from the National Natural Science Foundation (Grant No. 11925408, 11921004 and 12188101), the Ministry of Science and Technology of China (Grant No. 2018YFA0305700), the Chinese Academy of Sciences (Grant No. XDB33000000), the Informatization Plan of Chinese Academy of Sciences (Grant No. CAS-WX2021SF-0102), the K. C. Wong Education Foundation (GJTD-2018-01), the Beijing Natural Science Foundation (Z180008), and the Beijing Municipal Science and Technology Commission (Z191100007219013). X.D. acknowledges financial support from the Hong Kong Research Grants Council (Project No. GRF16300918 and No. 16309020).
\end{acknowledgments}

%


\end{document}


\title{Supplementary Materials: Topological States in Chevrel Phase Materials from First-principle Calculations}

\author{Shuai Zhang}
\affiliation{Beijing National Laboratory for Condensed Matter Physics, and Institute of Physics, Chinese Academy of Sciences, Beijing 100190, China}
\affiliation{University of Chinese Academy of Sciences, Beijing 100049, China}

\author{Shiyu Peng}
\affiliation{Beijing National Laboratory for Condensed Matter Physics, and Institute of Physics, Chinese Academy of Sciences, Beijing 100190, China}
\affiliation{University of Chinese Academy of Sciences, Beijing 100049, China}

\author{Xi Dai}
\affiliation{Department of Physics, Hong Kong University of Science and Technology, Clear Water Bay, Hong Kong}

\author{Hongming Weng}
\email{hmweng@iphy.ac.cn}
\affiliation{Beijing National Laboratory for Condensed Matter Physics, and Institute of Physics, Chinese Academy of Sciences, Beijing 100190, China}
\affiliation{University of Chinese Academy of Sciences, Beijing 100049, China}
\affiliation{Songshan Lake Materials Laboratory, Dongguan, Guangdong 523808, China}

\maketitle
\newpage

\section{The procedure of band renormalization}
Suppose our aim is to get a global gap between the $N$ lowest bands and the $(N+1)$-th band (i.e., a global gap in the $N$ occupation case) at the reference energy $\varepsilon_0$. 
We define $m_{k}=(E_{N+1,k} + E_{N,k})/2$ as the middle point of the $N$- and the $(N+1)$-th bands at $k$, where $E_{N,k}$ is the energy of the $N$-th band.
Then, we define $\delta_{k}=\varepsilon_0-m_{k}$. 
After shifting the energy of all the bands at $k$ with $\delta_{k}$, we can get a global gap at the reference energy $\varepsilon_0$. For convenience, we call this procedure ``renormalization''. There is no band inversion at each $k$ point during this procedure. Thus,  this kind of renormalization does not affect the band topology.

\section{The parity distribution, WCCs and surface states of $\BaMoS$}

\begin{figure}[H]
      \centering
      \includegraphics[width=1.0\linewidth]{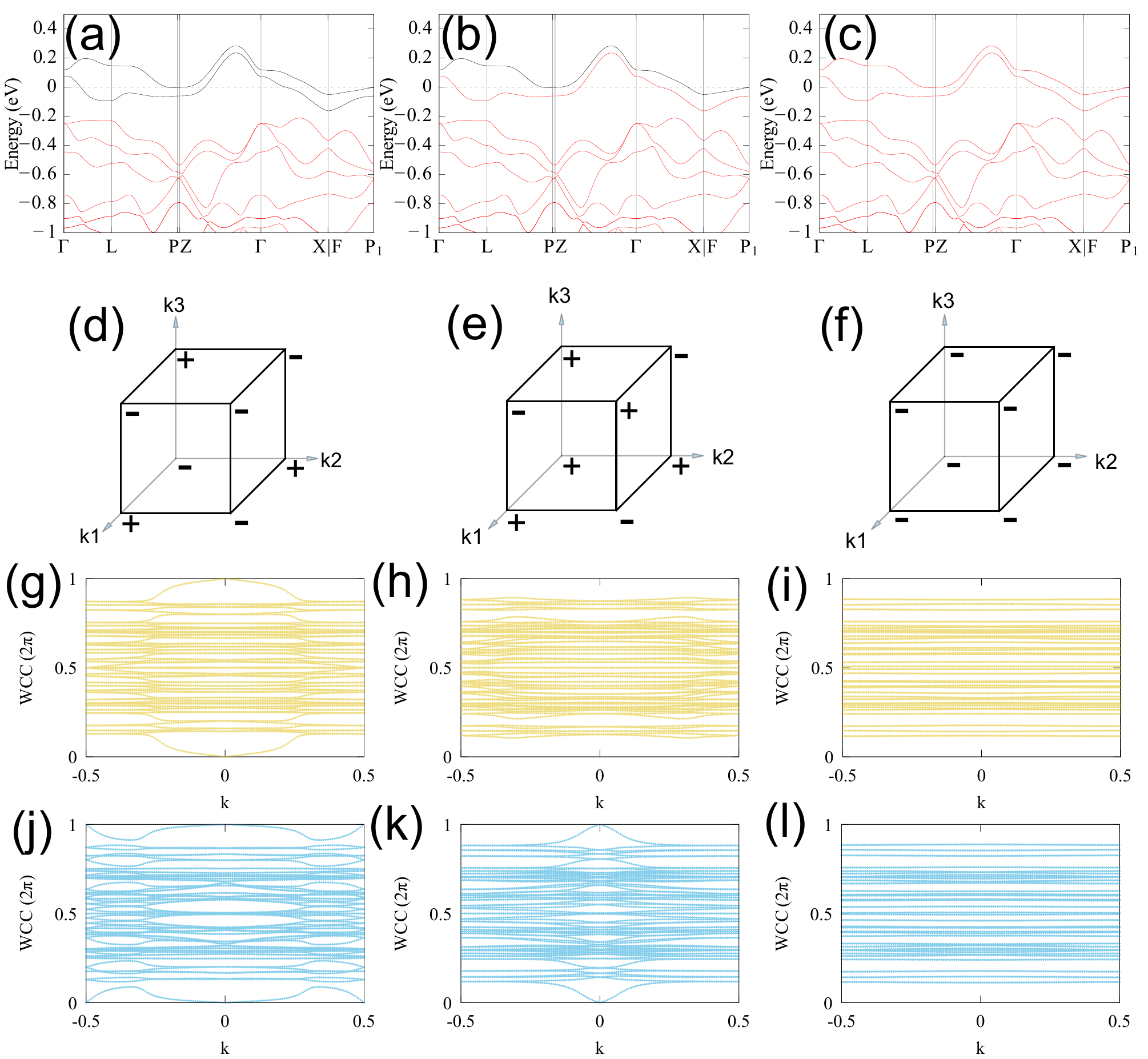}
      \caption{
        The band structures, parities, WCCs of $R\bar{3}$ phase $\BaMoS$ in different occupation cases with SOC.
        Here, each column belongs to the same occupation case.
        In (a), (b) and (c), the red bands represent the occupied ones, which show the $N-2$ occupation, $N$ occupation, and $N+2$ occupation cases intuitively.
        In (d), (e) and (f), ``-'' represents the product of  parity eigenvalues of all occupied Kramers degenerate pairs at this TRIM to be odd, and ``+'' represents even. 
       (g), (h), (i) show the evolution of WCCs on the $k_3=\pi$ plane, and Fig. (j), (k), (l) show the evolution of WCCs on the $k_3=0$ plane. Here, the WCCs integral direction is $k_1$ and WCCs evolve along $k_2$.
      }
      \label{fig:BaMoS_parity_wcc_148}
    \end{figure}

In Fig.~\ref{fig:BaMoS_parity_wcc_148}, we show the parity distribution and WCCs of $R\bar{3}$ phase $\BaMoS$ with SOC in the $N-2$ occupation, $N$ occupation and $N+2$ occupation cases, respectively.
These results indicate that the $\mathbb{Z}_2$ indices of $R\bar{3}$ phase $\BaMoS$ are $(1;0,0,0)$ in the $N$ occupation case and $(1;1,1,1)$ in the $N-2$ occupation case. 
Fig.~\ref{fig:BaMoS_parity_wcc_148}c, ~\ref{fig:BaMoS_parity_wcc_148}f, ~\ref{fig:BaMoS_parity_wcc_148}i and ~\ref{fig:BaMoS_parity_wcc_148}l demonstrate that the the lowest $N+2$ bands are topologically trivial, as we mentioned in the main text.

\begin{figure}[H]
      \centering
      \includegraphics[width=1.0\linewidth]{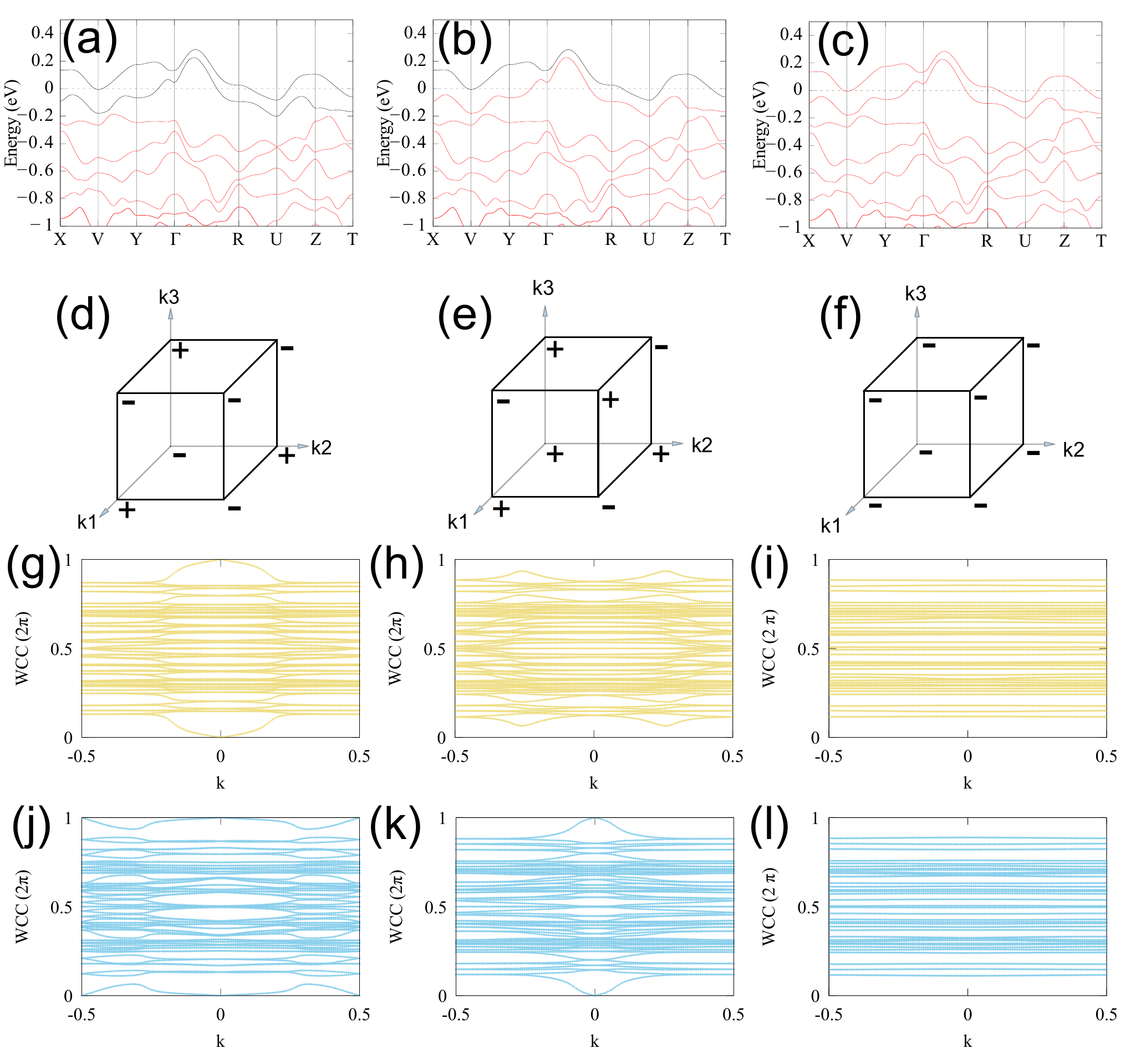}
      \caption{
        The band structures, parities, WCCs of $P\bar{1}$ phase $\BaMoS$ in different occupation cases with SOC.
        Here, each column belongs to the same occupation case.
        In  (a), (b) and (c), the red bands represent the occupied ones, which show the $N-2$ occupation, $N$ occupation, and $N+2$ occupation cases intuitively.
        In  (d), (e) and (f), ``-'' represents the product of  parity eigenvalues of all occupied Kramers degenerate pairs at this TRIM to be odd, and ``+'' represents even. 
         (g), (h), (i) show the evolution of WCCs on the $k_3=\pi$ plane, and  (j), (k), (l) show the evolution of  WCCs on the $k_3=0$ plane. Here, the WCCs integral direction is $k_1$ and WCCs evolves along $k_2$.
      }
      \label{fig:BaMoS_parity_wcc_2}
    \end{figure}

In Fig.~\ref{fig:BaMoS_parity_wcc_2}, we show the parity distribution and WCCs of $P\bar{1}$ phase $\BaMoS$ with SOC in the $N-2$ occupation, $N$ occupation and $N+2$ occupation cases, respectively.
These results indicate that the $\mathbb{Z}_2$ indices of $P\bar{1}$ phase $\BaMoS$ are the same as the ones of $R\bar{3}$ phase $\BaMoS$ in both of the $N-2$ and $N$ occupation cases.
Fig.~\ref{fig:BaMoS_parity_wcc_2}c, ~\ref{fig:BaMoS_parity_wcc_2}f, ~\ref{fig:BaMoS_parity_wcc_2}i and ~\ref{fig:BaMoS_parity_wcc_2}l demonstrate that the the lowest $N+2$ bands are topologically trivial.

\begin{figure}[H]
      \centering
      \includegraphics[width=1.0\linewidth]{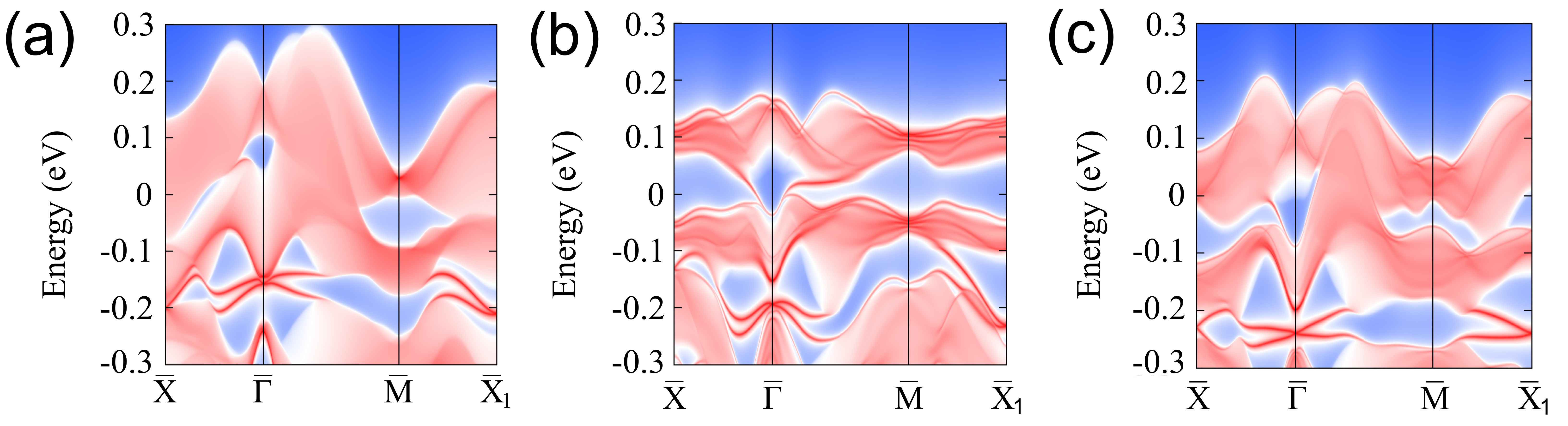}
      \caption{
           The $(001)$ surface states of $P\bar{1}$ phase $\BaMoS$ with SOC. 
           (a) The surface states calculated with the original band structure.
           (b) The surface states calculated with the renormalized band structure, which has a global gap in the $N$ occupation case.
           (c) The surface states calculated with the renormalized band structure, which has a global gap in the $N-2$ occupation case.
      }
      \label{fig:BaMoS_surface_P_1}
    \end{figure}

The $(001)$ surface states of $P\bar{1}$ phase $\BaMoS$ with SOC are shown in Fig.~\ref{fig:BaMoS_surface_P_1}a.
Because of the absence of global gap in both of the $N$ and $N-2$ occupation cases, the surface Dirac cone is mixed with the bulk states.
We calculated the renormalized band structures and the corresponding surface states with 0.02 eV and 0.01 eV onsite energy correction to the atoms in the surface region for the $N$ occupation and $N-2$ occupation cases, respectively.
As shown in Fig.~\ref{fig:BaMoS_surface_P_1}b and ~\ref{fig:BaMoS_surface_P_1}c,  there is a global gap in bulk states and clear surface Dirac cones in both cases. 
There is only one surface Dirac cone at $\bar{\Gamma}$ in the $N$ occupation case and three surface Dirac cones at $\bar{\rm X}$, $\bar{\Gamma}$ and $\bar{\rm X}_1$ in the $N-2$ occupation case.
The number and the distribution of surface Dirac cones are consistent with the parity distribution.

\section{The band structure, parity distribution, WCCs and surface states of $\SrMoS$}

\begin{figure}[H]
      \centering
      \includegraphics[width=0.9\linewidth]{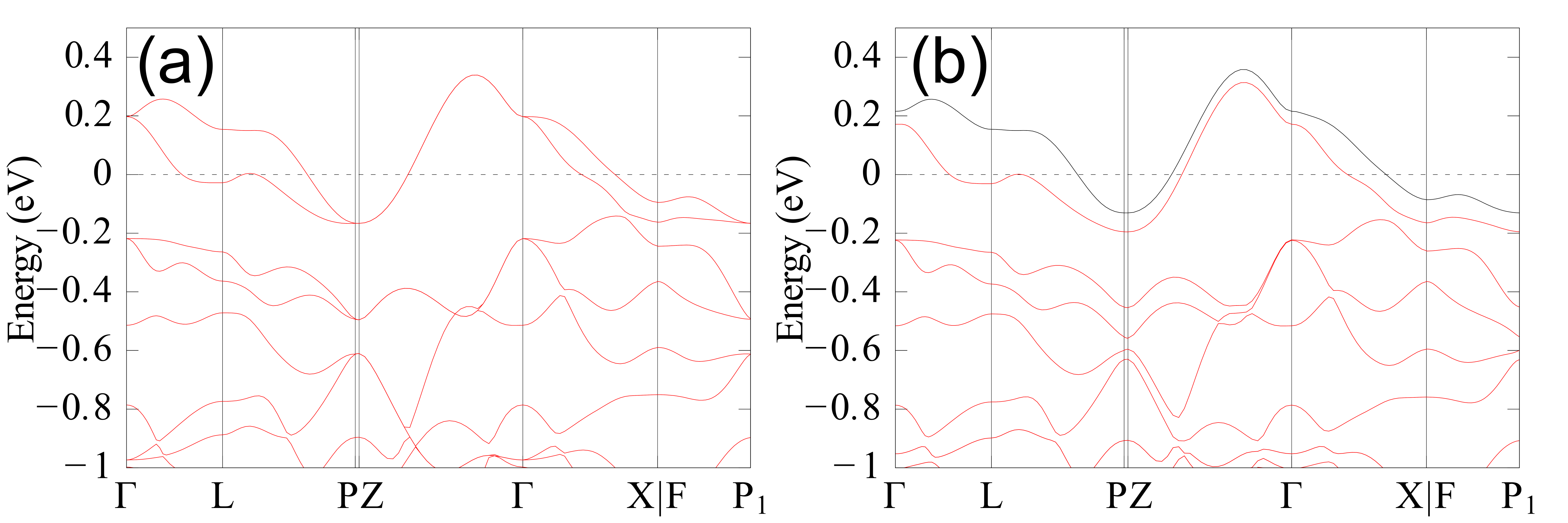}
      \caption{
      The band structures of $R\bar{3}$ phase $\SrMoS$ without SOC (a) and with SOC (b). 
      In (b), the red bands are the $N$ occupied ones, where $N$ represents the number of valence electrons of $\BaMoS$.
      }
      \label{fig:SrMoS_band_148}
    \end{figure}

Fig.~\ref{fig:SrMoS_band_148} shows the band structures of $R\bar{3}$ phase $\SrMoS$, which are basically the same as the ones of $R\bar{3}$ phase $\BaMoS$.

\begin{figure}[H]
      \centering
      \includegraphics[width=1.0\linewidth]{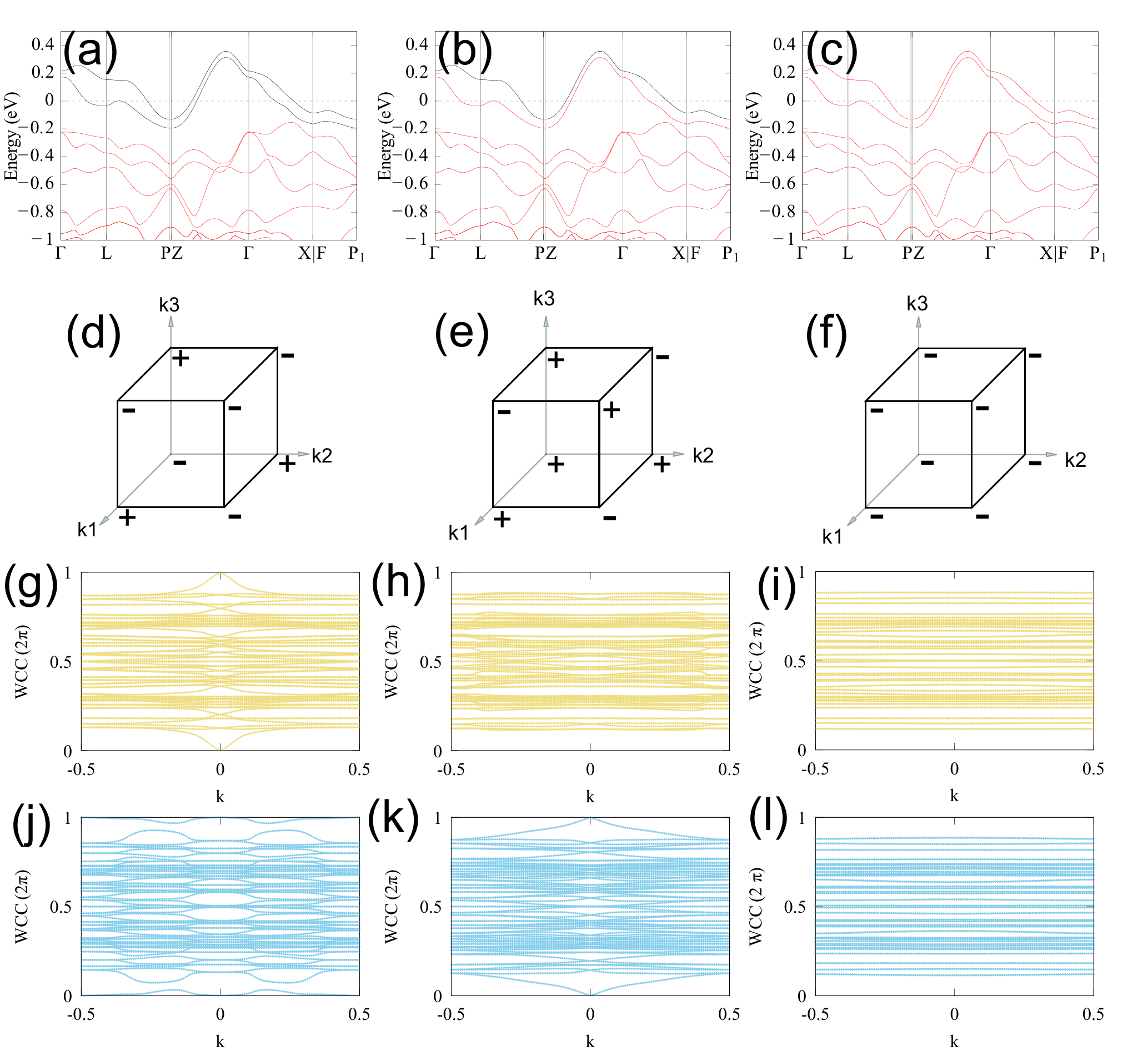}
      \caption{
        The band structures, parities, WCC of $R\bar{3}$ $\SrMoS$ in different occupation cases with SOC.
        Here, each column belongs to the same occupation case.
        In (a), (b) and (c), the red bands represent the occupied ones, which show the $N-2$ occupation, $N$ occupation, and $N+2$ occupation cases intuitively.
        In (d), (e) and (f), ``-'' represents the product of  parity eigenvalues of all occupied Kramers degenerate pairs at this TRIM to be odd, and ``+'' represents even. 
        (g), (h), (i) show the evolution of WCCs on the $k_3=\pi$ plane, and (j), (k), (l) show the evolution of WCC on the $k_3=0$ plane. Here, the WCCs integral direction is $k_1$ and WCCs evolve along $k_2$.
      }
      \label{fig:SrMoS_parity_wcc_148}
    \end{figure}

In Fig.~\ref{fig:SrMoS_parity_wcc_148}, we show the parity distribution and WCCs of $R\bar{3}$ phase $\SrMoS$ with SOC in the $N-2$ occupation, $N$ occupation and $N+2$ occupation cases, respectively. 
These results indicate that the $\mathbb{Z}_2$ indices of $R\bar{3}$ phase $\SrMoS$ are $(1;0,0,0)$ in the $N$ occupation case and $(1;1,1,1)$ in the $N-2$ occupation case. Fig.~\ref{fig:SrMoS_parity_wcc_148}c, ~\ref{fig:SrMoS_parity_wcc_148}f, ~\ref{fig:SrMoS_parity_wcc_148}i and ~\ref{fig:SrMoS_parity_wcc_148}l demonstrate that the the lowest $N+2$ bands are topologically trivial

\begin{figure}[H]
      \centering
      \includegraphics[width=1.0\linewidth]{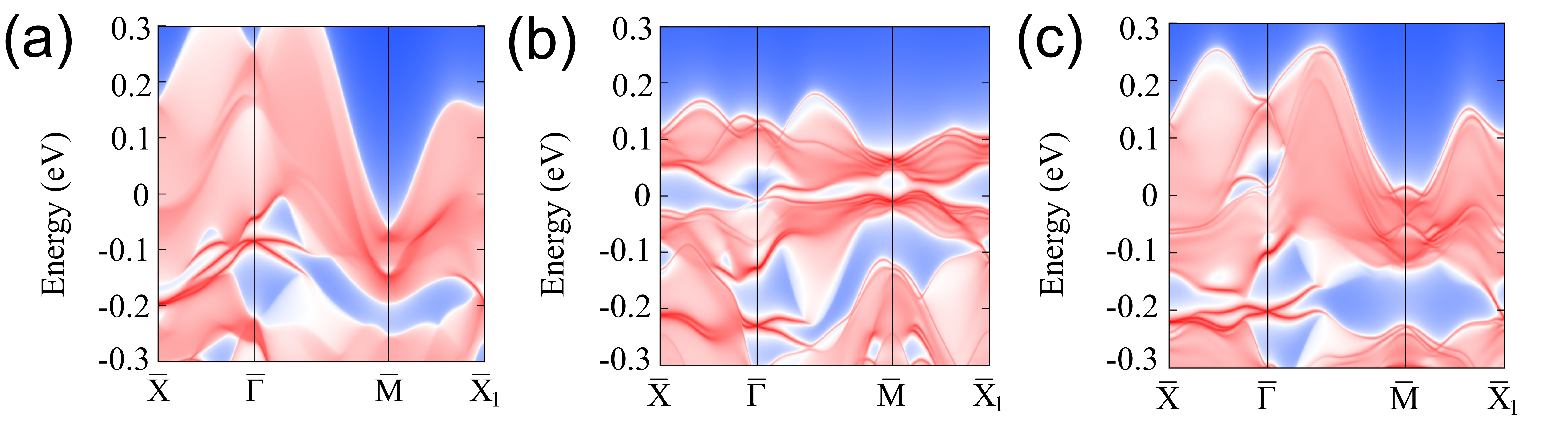}
      \caption{
           The $(001)$ surface states of $R\bar{3}$ phase $\SrMoS$ with SOC. 
           (a) The surface states calculated with the original band structure.
           (b) The surface states calculated with the renormalized band structure, which has a global gap in the $N$ occupation case.
           (c) The surface states calculated with the renormalized band structure, which has a global gap in the $N-2$ occupation case.
      }
      \label{fig:SrMoS_surface_148}
    \end{figure}

The $(001)$ surface states of $R\bar{3}$ phase $\SrMoS$ with SOC are shown in Fig.~\ref{fig:SrMoS_surface_148}a.
Because of the absence of global gap in both of the $N$ and $N-2$ occupation cases, the surface Dirac cone is mixed with the bulk states.
We calculated the renormalized band structures and the corresponding surface states with 0.02 eV onsite energy correction to the atoms in the surface region for the $N$ occupation and $N-2$ occupation cases, respectively.
As shown in Fig.~\ref{fig:SrMoS_surface_148}b and ~\ref{fig:SrMoS_surface_148}c, there is a global gap in bulk states and clear surface Dirac cones in both cases. 
There is only one surface Dirac cone at $\bar{\Gamma}$ in the $N$ occupation case and three surface Dirac cones at $\bar{\rm X}$, $\bar{\rm \Gamma}$ and $\bar{\rm X}_1$ in the $N-2$ occupation case.
The number and the distribution of surface Dirac cones are consistent with the parity distribution.

\begin{figure}[H]
      \centering
      \includegraphics[width=1.0\linewidth]{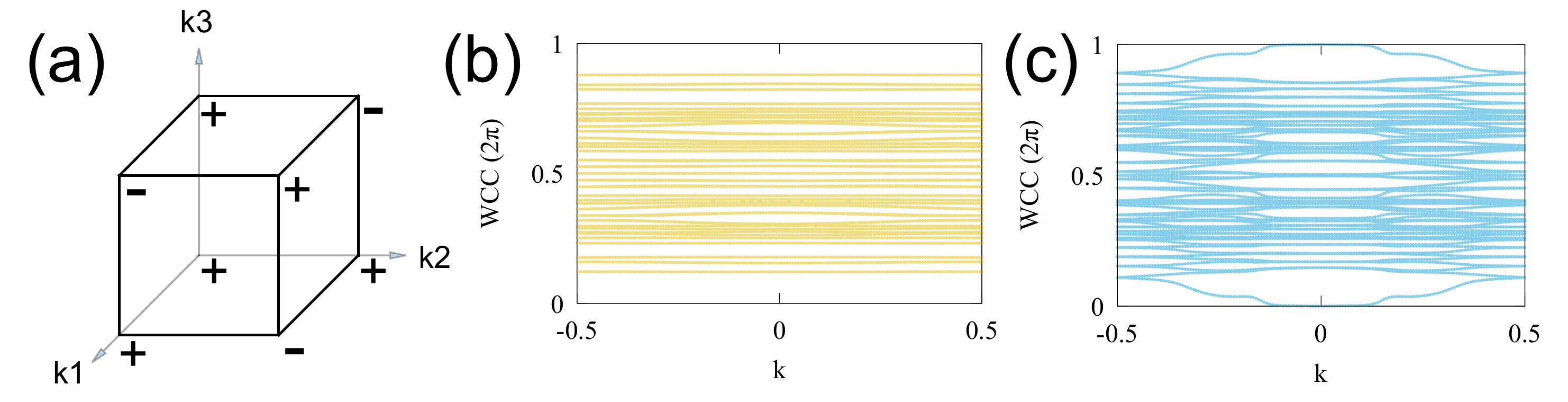}
      \caption{
        The information of the topological state of  $P\bar{1}$ phase $\SrMoS$ with SOC.
        (a) The parity distribution, where the ``+'' represents even and the ``-'' represents odd.  
        (b) and (c) show the evolution of WCCs on the $k_3=\pi$ plane and the $k_3=0$ plane, respectively.  Here, the WCCs integral direction is $k_1$ and WCCs evolve along $k_2$. 
      }
      \label{fig:SrMoS_topo_2}
    \end{figure}

The parity distribution and WCCs of $P\bar{1}$ phase $\SrMoS$ with SOC in the $N$ occupation case are shown in Fig.~\ref{fig:SrMoS_topo_2}.
These results show that the $\mathbb{Z}_2$ indices of $P\bar{1}$ phase $\SrMoS$ are $(1;0,0,0)$ in the $N$ occupation case.

\begin{figure}[H]
      \centering
      \includegraphics[width=1.0\linewidth]{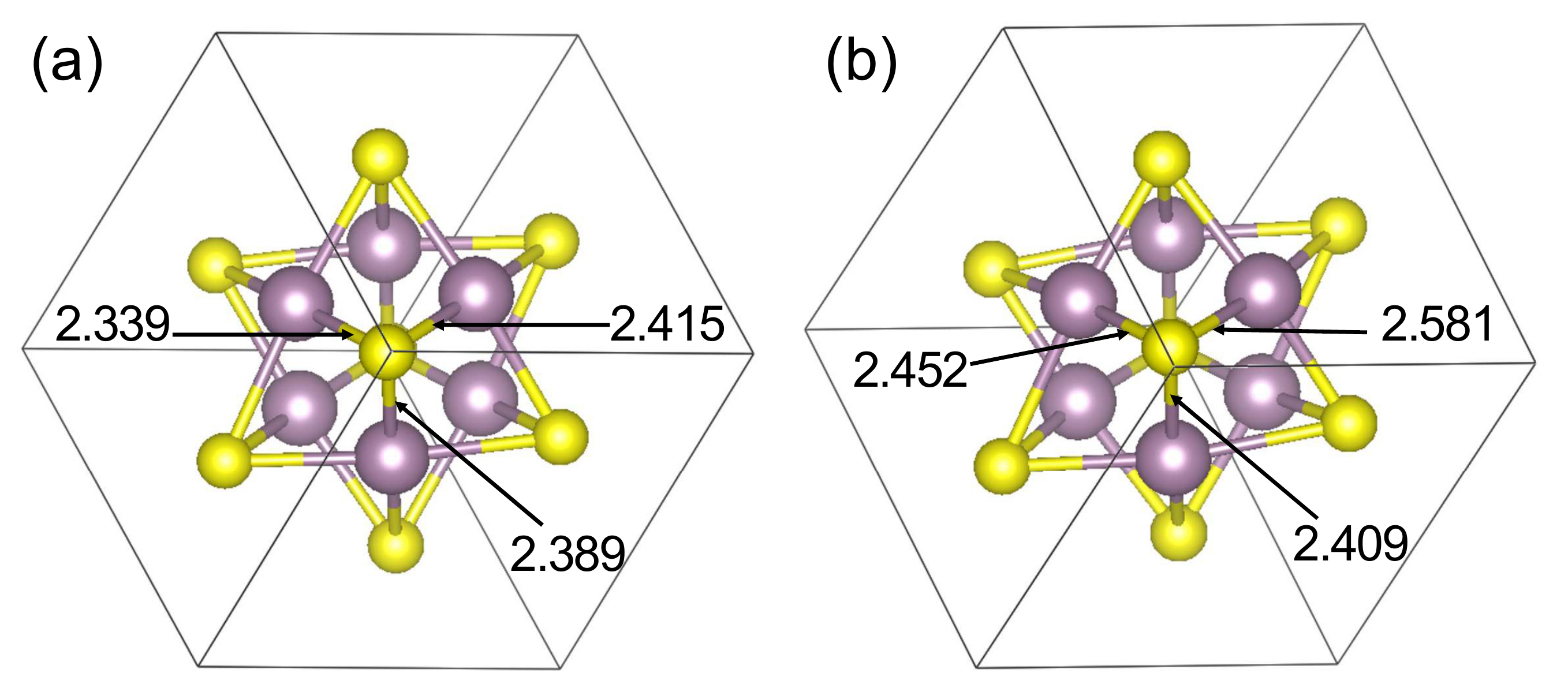}
      \caption{
        The $\MoS$ clusters of $\BaMoS$ (a) and $\SrMoS$ (b) in the $P\bar{1}$ phase and the three bond lengths (in unit \AA) is labeled.
        Here we omitted the Ba (Sr) atoms for clearness.
      }
      \label{fig:deformation_from_C3}
    \end{figure}

The three Mo-S bond lengths in $P\bar{1}$ phase $\BaMoS$ and $\SrMoS$ are shown in Fig.~\ref{fig:deformation_from_C3} to compare the difference in their crystal structures.

\section{The parity distribution, WCCs and surface states of $\MoS$}

\begin{figure}[H]
      \centering
      \includegraphics[width=1.0\linewidth]{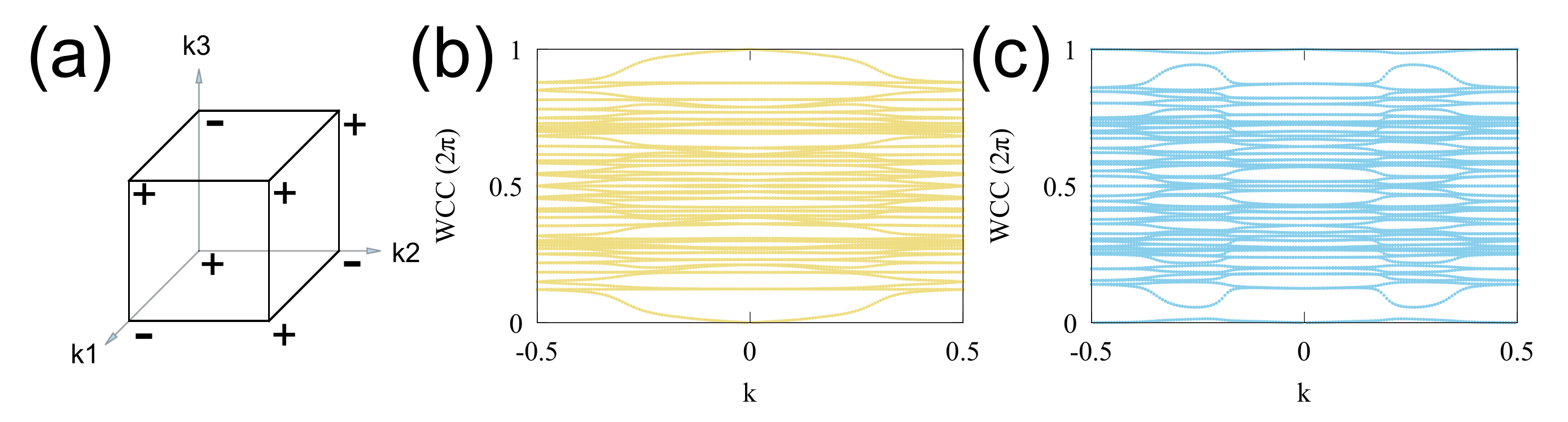}
      \caption{
        The information of the topological states of $R\bar{3}$ phase $\MoS$ with SOC.
        (a) The parity distribution, where ``+'' represents even and ``-'' represents odd. 
        (b) and (c) show the evolution of WCCs on the $k_3=\pi$ plane and the $k_3=0$ plane, respectively. Here, the WCCs integral direction is $k_1$ and WCCs evolve along $k_2$.
      }
      \label{fig:MoS_topo}
    \end{figure}

In Fig.~\ref{fig:MoS_topo}, we show the parity distribution and WCCs of $R\bar{3}$ phase $\MoS$ with SOC. 
These results show that the $\mathbb{Z}_2$ indices of $R\bar{3}$ phase $\MoS$ are $(1;0,0,0)$, which are the same as the ones of $R\bar{3}$ phase $\BaMoS$ in the $N-2$ occupation case.

\begin{figure}[H]
      \centering
      \includegraphics[width=1.0\linewidth]{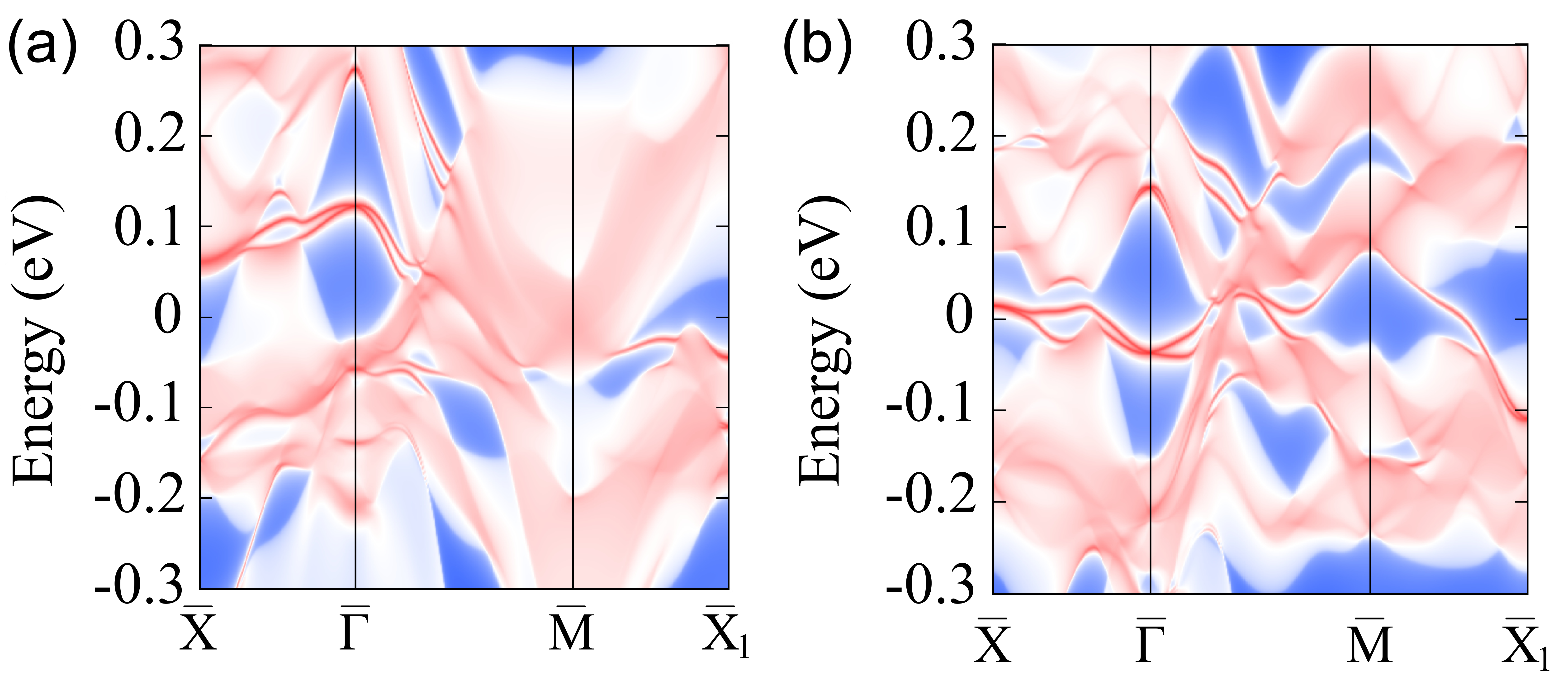}
      \caption{
      The $(001)$ surface states of $R\bar{3}$ phase $\MoS$ with SOC.
         (a) The surface states calculated with the original band structure.
         (b) The surface states calculated with the renormalized band structure with a global gap.
      }
      \label{fig:MoS_surf}
    \end{figure}

The $(001)$ surface states of $R\bar{3}$ phase $\MoS$ are shown in Fig.~\ref{fig:MoS_surf}a. 
Because of the absence of global gap, the surface Dirac cone is mixed with the bulk states.
We calculated the renormalized band structure, which has a global gap, and the corresponding surface states, as shown in Fig.~\ref{fig:MoS_surf}b.
There are three surface Dirac cones at $\bar{\rm X}$, $\bar{\Gamma}$ and $\bar{\rm X}_1$.
The number and the distribution of surface Dirac cones are consistent with the parity distribution.

\begin{figure}[H]
      \centering
      \includegraphics[width=0.5\linewidth]{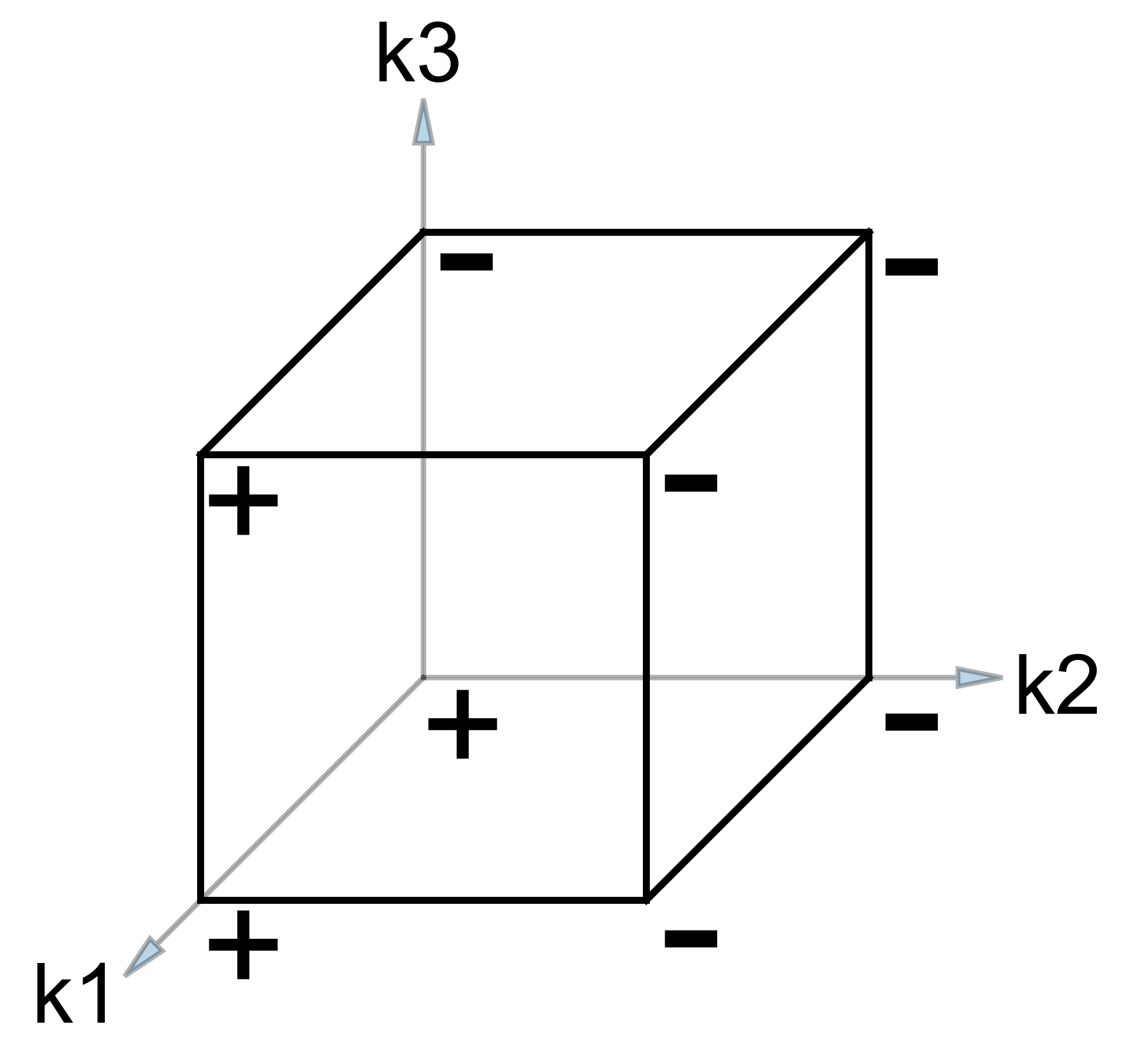}
      \caption{
        The parity distribution of $P\bar{1}$ phase $\MoS$ with SOC, where ``+'' represents even, and ``-'' represents   odd. 
      }
      \label{fig:MoS_parity_2}
    \end{figure}

The parity distribution with SOC of $P\bar{1}$ phase $\MoS$ is shown in Fig.~\ref{fig:MoS_parity_2}, which indicates the $\mathbb{Z}_2$ indices of $P\bar{1}$ phase $\MoS$ are $(1;0,0,1)$.


\section{The crystal structures of $\BaMoS$, $\SrMoS$, $\MoS$ and $\EuMoS$}

We list the crystal structures used in calculations in Tab.~\ref{tab:Ba_struct}-~\ref{tab:Eu_struct}.

\begin{table}[H]
  \centering
  \caption{The crystal structures of $\BaMoS$ in the $R\bar{3}$ phase (ICDD PDF: 01-088-1169)~\cite{BaR3_P1_structure_1986} and the $P\bar{1}$ phase (ICDD PDF: 01-088-1170)~\cite{BaR3_P1_structure_1986}.}
    \begin{tabular}{lrrrrrrr}
          &       &       &       &       &       &       &  \\
    \hline
    \multicolumn{8}{c}{$\BaMoS$} \\
    \hline
          & \multicolumn{3}{c}{$R\bar{3}$} &       & \multicolumn{3}{c}{$P\bar{1}$} \\
    \hline
    Lattice vectors (\AA) &       &       &       &       &       &       &  \\
    a     & 4.64870  & 2.68393  & 3.92447  &       & -6.56850  & 0.00000  & 0.00000  \\
    b     & -4.64870  & 2.68393  & 3.92447  &       & -0.11946  & -0.12586  & -6.69535  \\
    c     & 0.00000  & -5.36786  & 3.92447  &       & -0.24048  & -6.65015  & 0.00000  \\
    \hline
    Atom positions (fraction) &       &       &       &       &       &       &  \\
    Ba    & 0.00000  & 0.00000  & 0.00000  &       & 0.00000  & 0.00000  & 0.00000  \\
    Mo    & 0.76541  & 0.43444  & 0.58245  &       & -0.58084  & -0.76250  & -0.43557  \\
    Mo    & 0.23459  & 0.56556  & 0.41755  &       & -0.41916  & -0.23750  & -0.56443  \\
    Mo    & 0.56556  & 0.41755  & 0.23459  &       & -0.43472  & -0.58235  & -0.76695  \\
    Mo    & 0.58245  & 0.76541  & 0.43444  &       & -0.56528  & -0.41765  & -0.23305  \\
    Mo    & 0.41755  & 0.23459  & 0.56556  &       & -0.76736  & -0.43322  & -0.58331  \\
    Mo    & 0.43444  & 0.58245  & 0.76541  &       & -0.23264  & -0.56678  & -0.41669  \\
    S     & 0.60969  & 0.26402  & 0.87703  &       & -0.26310  & -0.87600  & -0.60760  \\
    S     & 0.39031  & 0.73598  & 0.12297  &       & -0.73690  & -0.12400  & -0.39240  \\
    S     & 0.73598  & 0.12297  & 0.39031  &       & -0.74760  & -0.74440  & -0.75310  \\
    S     & 0.87703  & 0.60969  & 0.26402  &       & -0.25240  & -0.25560  & -0.24690  \\
    S     & 0.12297  & 0.39031  & 0.73598  &       & -0.87530  & -0.60850  & -0.26410  \\
    S     & 0.26402  & 0.87703  & 0.60969  &       & -0.12470  & -0.39150  & -0.73590  \\
    S     & 0.74850  & 0.74850  & 0.74850  &       & -0.61220  & -0.26240  & -0.87760  \\
    S     & 0.25150  & 0.25150  & 0.25150  &       & -0.38780  & -0.73760  & -0.12240  \\
    \hline
    \end{tabular}%
  \label{tab:Ba_struct}%
\end{table}%

\begin{table}[H]
  \centering
  \caption{The crystal structures of $\SrMoS$ in the $R\bar{3}$ phase (ICDD PDF: 01-085-9202)~\cite{Sr_R3_structure} and the $P\bar{1}$ phase (ICDD PDF: 04-001-6572)~\cite{Sr_P1_structure}. }
    \begin{tabular}{lrrrrrrr}
    \hline
    \multicolumn{8}{c}{$\SrMoS$} \\
    \hline
          & \multicolumn{3}{c}{$R\bar{3}$} &       & \multicolumn{3}{c}{$P\bar{1}$} \\
    \hline
    Lattice vectors (\AA) &       &       &       &       &       &       &  \\
    a     & 4.59480  & 2.65281  & 3.84773  &       & -6.48100  & 0.00000  & 0.00000  \\
    b     & -4.59480  & 2.65281  & 3.84773  &       & -0.08030  & -0.08447  & -6.60997  \\
    c     & 0.00000  & -5.30562  & 3.84773  &       & -0.20999  & -6.56864  & 0.00000  \\
    \hline
    Atom positions (fraction) &       &       &       &       &       &       &  \\
    Sr    & 0.00000  & 0.00000  & 0.00000  &       & 0.00000  & 0.00000  & 0.00000  \\
    Mo    & 0.22900  & 0.41799  & 0.56300  &       & -0.23100  & -0.41700  & -0.56400  \\
    Mo    & 0.77100  & 0.58201  & 0.43700  &       & -0.76900  & -0.58300  & -0.43600  \\
    Mo    & 0.56300  & 0.22900  & 0.41799  &       & -0.41800  & -0.56800  & -0.23200  \\
    Mo    & 0.43700  & 0.77100  & 0.58201  &       & -0.58200  & -0.43200  & -0.76800  \\
    Mo    & 0.41799  & 0.56300  & 0.22900  &       & -0.56700  & -0.22600  & -0.41800  \\
    Mo    & 0.58201  & 0.43700  & 0.77100  &       & -0.43300  & -0.77400  & -0.58200  \\
    S     & 0.38400  & 0.12801  & 0.74100  &       & -0.12600  & -0.76500  & -0.38500  \\
    S     & 0.61600  & 0.87199  & 0.25900  &       & -0.87400  & -0.23500  & -0.61500  \\
    S     & 0.74100  & 0.38400  & 0.12801  &       & -0.23200  & -0.22500  & -0.25500  \\
    S     & 0.25900  & 0.61600  & 0.87199  &       & -0.76800  & -0.77500  & -0.74500  \\
    S     & 0.12801  & 0.74100  & 0.38400  &       & -0.37300  & -0.12300  & -0.71800  \\
    S     & 0.87199  & 0.25900  & 0.61600  &       & -0.62700  & -0.87700  & -0.28200  \\
    S     & 0.24700  & 0.24700  & 0.24700  &       & -0.74800  & -0.35900  & -0.14400  \\
    S     & 0.75300  & 0.75300  & 0.75300  &       & -0.25200  & -0.64100  & -0.85600  \\
    \hline
    \end{tabular}%
  \label{tab:Sr_struct}%
\end{table}%

\begin{table}[H]
  \centering
  \caption{Information of the crystal structures of $\MoS$ in the $R\bar{3}$ phase (SpringerMaterials Dataset ID: sd\_0312550)~\cite{MoS_springer,MoS_structure} and the $P\bar{1}$ phase (ICDD PDF: 04-020-4062)~\cite{MoS2_struct}.}
    \begin{tabular}{lrrrrrrr}
    \hline
    \multicolumn{8}{c}{$\MoS$} \\
    \hline
          & \multicolumn{3}{c}{$R\bar{3}$} &       & \multicolumn{3}{c}{$P\bar{1}$} \\
    \hline
    Lattice vectors (\AA) &       &       &       &       &       &       &  \\
    a     & 4.59460  & 2.65269  & 3.62933  &       & 1.59392  & -6.41288  & 0.00000  \\
    b     & -4.59460  & 2.65269  & 3.62933  &       & 6.36400  & 0.00000  & 0.00000  \\
    c     & 0.00000  & -5.30539  & 3.62933  &       & 3.19332  & -2.47343  & 5.49467  \\
    \hline
    Atom positions (fraction) &       &       &       &       &       &       &  \\
    Mo    & 0.77696  & 0.58703  & 0.45392  &       & -0.00147  & -0.22401  & 0.50321  \\
    Mo    & 0.22304  & 0.41297  & 0.54608  &       & -0.99853  & 0.22401  & 0.49679  \\
    Mo    & 0.45392  & 0.77696  & 0.58703  &       & -0.48276  & 0.27235  & 0.18129  \\
    Mo    & 0.54608  & 0.22304  & 0.41297  &       & -0.51724  & -0.27235  & 0.81871  \\
    Mo    & 0.58703  & 0.45392  & 0.77696  &       & -0.48563  & 0.78343  & 0.19147  \\
    Mo    & 0.41297  & 0.54608  & 0.22304  &       & -0.51437  & -0.78343  & 0.80853  \\
    S     & 0.62083  & 0.87260  & 0.25530  &       & -0.21270  & -0.41750  & 0.94820  \\
    S     & 0.37917  & 0.12740  & 0.74470  &       & -0.78730  & 0.41750  & 0.05180  \\
    S     & 0.25530  & 0.62083  & 0.87260  &       & -0.26930  & -0.41800  & 0.47220  \\
    S     & 0.74470  & 0.37917  & 0.12740  &       & -0.73070  & 0.41800  & 0.52780  \\
    S     & 0.87260  & 0.25530  & 0.62083  &       & -0.79460  & 0.92020  & 0.06350  \\
    S     & 0.12740  & 0.74470  & 0.37917  &       & -0.20540  & -0.92020  & 0.93650  \\
    S     & 0.78627  & 0.78627  & 0.78627  &       & -0.71490  & -0.08660  & 0.53200  \\
    S     & 0.21373  & 0.21373  & 0.21373  &       & -0.28510  & 0.08660  & 0.46800  \\    
    \hline
    \end{tabular}%
  \label{tab:MoS_struct}%
\end{table}%

\begin{table}[H]
  \centering
  \caption{Information of the crystal structures of $\EuMoS$ in the $R\bar{3}$ phase (ICDD PDF: 00-048-1791)~\citep{Eu_R3_structure_3} and the $P\bar{1}$ phase (ICDD PDF: 01-079-0877)~\citep{Eu_P1_structure}.}
    \begin{tabular}{lrrrrrrr}
    \hline
    \multicolumn{8}{c}{$\EuMoS$} \\
    \hline
          & \multicolumn{3}{c}{$R\bar{3}$} &       & \multicolumn{3}{c}{$P\bar{1}$} \\
    \hline
    Lattice vectors (\AA) &       &       &       &       &       &       &  \\
    a     & 4.58870  & 2.64929  & 3.85410  &       & -0.08492  & -0.08293  & -6.59193  \\
    b     & -4.58870  & 2.64929  & 3.85410  &       & -0.21025  & -6.56263  & 0.00000  \\
    c     & 0.00000  & -5.29857  & 3.85410  &       & -6.48200  & 0.00000  & 0.00000  \\
    \hline
    Atom positions (fraction) &       &       &       &       &       &       &  \\
    Eu    & 0.00000  & 0.00000  & 0.00000  &       & 0.00000  & 0.00000  & 0.00000  \\
    Mo    & 0.43803  & 0.77274  & 0.58377  &       & -0.43890  & -0.58160  & -0.76700  \\
    Mo    & 0.56197  & 0.22726  & 0.41623  &       & -0.56110  & -0.41840  & -0.23300  \\
    Mo    & 0.22726  & 0.41623  & 0.56197  &       & -0.76950  & -0.44380  & -0.58360  \\
    Mo    & 0.58377  & 0.43803  & 0.77274  &       & -0.23050  & -0.55620  & -0.41640  \\
    Mo    & 0.41623  & 0.56197  & 0.22726  &       & -0.58550  & -0.78070  & -0.43850  \\
    Mo    & 0.77274  & 0.58377  & 0.43803  &       & -0.41450  & -0.21930  & -0.56150  \\
    S     & 0.87471  & 0.25789  & 0.61699  &       & -0.60200  & -0.23600  & -0.85400  \\
    S     & 0.12529  & 0.74211  & 0.38301  &       & -0.39800  & -0.76400  & -0.14600  \\
    S     & 0.74211  & 0.38301  & 0.12529  &       & -0.75400  & -0.76500  & -0.75200  \\
    S     & 0.61699  & 0.87471  & 0.25789  &       & -0.24600  & -0.23500  & -0.24800  \\
    S     & 0.38301  & 0.12529  & 0.74211  &       & -0.26200  & -0.84700  & -0.62000  \\
    S     & 0.25789  & 0.61699  & 0.87471  &       & -0.73800  & -0.15300  & -0.38000  \\
    S     & 0.75637  & 0.75637  & 0.75637  &       & -0.88900  & -0.62300  & -0.26000  \\
    S     & 0.24363  & 0.24363  & 0.24363  &       & -0.11100  & -0.37700  & -0.74000  \\
    \hline
    \end{tabular}%
  \label{tab:Eu_struct}%
\end{table}%

%
